 \documentclass[preprint,review,12pt]{elsarticle}

\usepackage{amssymb}
\usepackage{amsmath}

\usepackage{graphicx}%
\usepackage{multirow}%
\usepackage{amsmath,amssymb,amsfonts}%
\usepackage{amsthm}%
\usepackage{mathrsfs}%
\usepackage[title]{appendix}%
\usepackage{xcolor}%
\usepackage{textcomp}%
\usepackage{manyfoot}%
\usepackage{booktabs}%
\usepackage{algorithm}%
\usepackage{algorithmicx}%
\usepackage{algpseudocode}%
\usepackage{listings}%
\usepackage{bm}
\usepackage{subcaption}
\usepackage{caption}
\usepackage{pdflscape}
\usepackage{soul}

\journal{Acta Astronautica}

\begin{document}

\begin{frontmatter}



\title{Pulsar Selection Criteria and Performance Evaluation of Autonomous X-ray Pulsar Navigation Systems}

 \author[label1,label2,label3]{Sui Chen}
  \author[label2,label3]{Emilie Parent}
   \author[label2,label3]{Nanda Rea}
 \author[label1]{Francesco Topputo}
 \affiliation[label1]{organization={Department of Aerospace Science and Technology, Politecnico di Milano},
             addressline={Via La Masa 34},
             city={Milan},
             postcode={20156},
             country={Italy}}

 \affiliation[label2]{organization={Institute of Space Sciences (ICE-CSIC), Campus UAB},
             addressline={Carrer de Can Magrans s/n, Cerdanyola del Vallès},
             city={Barcelona},
             postcode={08193},
             country={Spain}}

 \affiliation[label3]{organization={Institut d’Estudis Espacials de Catalunya (IEEC)},
             addressline={Carrer Gran Capità 2-4, },
             city={Barcelona},
             postcode={08034},
             country={Spain}}



\begin{abstract}
Current space missions primarily depend on Earth-based Guidance, Navigation, and Control (GNC) systems involving human-in-the-loop operations. X-ray pulsar-based navigation offers a promising alternative by using the very precise periodic X-ray emissions from pulsars for fully autonomous state estimation. This study presents a comprehensive analysis of pulsar selection criteria that significantly influence overall navigation performance. Observational data from the NICER mission is used to derive realistic estimates of measurement noise. Key mission-level constraints, including pulsed flux, pulsar visibility, geometric configuration, and long-term timing stability, are integrated into the pulsar selection process, addressing limitations of existing studies. An extended Kalman filter (EKF) is used for onboard spacecraft state estimation. The proposed system is evaluated in two scenarios:  a Low Earth Orbit (LEO) satellite at 600 km altitude and an interplanetary transfer from Earth to Jupiter. Simulation results show that including the Crab pulsar yields position errors below 7 km in LEO and 20 km during interplanetary transfer with an instrument effective area of 200~cm$^2$; however, the Crab's limited timing stability leads to filter divergence after 20 days without timing model updates. In contrast, more stable pulsars enable long-term autonomy but with reduced accuracy. These results highlight the trade-offs involved in pulsar selection for autonomous navigation and the need to balance competing objectives. Overall, this study demonstrates the feasibility of X-ray pulsar-based navigation and marks a key step towards fully autonomous spacecraft operations.
\end{abstract}




\end{frontmatter}



\section{Introduction}

The rapid progress in space technology, coupled with the increasing commercialisation of the space sector, has led to a sharp rise in the number and complexity of space exploration missions. Currently, most of these missions rely on ground-based guidance, navigation, and control (GNC) systems, which often require human oversight. Although such systems have proven effective and reliable, they are not without limitations. Communication latency, restricted real-time adaptability, and the growing demand placed on ground infrastructure, especially as more actors enter the space domain, pose significant operational challenges. As a result, the development of onboard autonomous navigation capabilities is gaining increasing attention, aiming to enhance mission resilience and reduce dependence on Earth-based support.

Among the various approaches being explored, X-ray pulsar navigation (XNAV) stands out as an attractive solution. This method uses the periodic timing signals emitted by pulsars, fast-spinning neutron stars with highly stable and predictable pulse profiles, as natural beacons for navigation. These features give pulsar-based navigation an extremely high achievable accuracy. The idea of using pulsars as navigation tools was first proposed in the 1970s by Downs \cite{downs1974interplanetary}, who investigated the potential of radio pulsars for interplanetary navigation. Across the electromagnetic spectrum, X-ray pulsars are particularly suitable for spacecraft navigation due to their predictable timing that is unaffected by interstellar dispersion, and the possibility of using compact, low-power detectors, in contrast to the larger, more power-intensive systems required for radio observations \cite{sheikh2006spacecraft}.


In 2016, the Chinese mission XPNAV-1 was launched to perform the first on-ground demonstration of pulsar-based navigation using data collected onboard. The experiment achieved an average navigation accuracy of 38.4~km using observations of the Crab pulsar alone \cite{huang2019pulsar}. In 2017, NASA's SEXTANT experiment was conducted in conjunction with the NICER mission onboard the International Space Station (ISS), marking the first real-time, fully autonomous XNAV demonstration in space. This system relied on sequential observations of multiple millisecond pulsars and maintained position errors below 10~km, with a filter $3\sigma$ covariance under 30~km \cite{mitchell2018sextant,winternitz2018sextant}. However, the SEXTANT experiment employed a large X-ray collecting area exceeding 1800~cm$^2$, which is impractical for small spacecraft platforms. Moreover, this demonstration was conducted in low Earth orbit (LEO), where well-established navigation technologies, particularly GPS, already provide highly accurate solutions at the metre level. In contrast, XNAV is more advantageous for deep-space and interplanetary missions, where GPS signals are unavailable and planetary features suitable for optical navigation may not be consistently observable.

Apart from on-ground and in-orbit demonstrations of XNAV, substantial research efforts have focused on refining algorithms and understanding key error sources. Early work by Sheikh \cite{sheikh2006spacecraft} provided the theoretical basis for later developments, after which a range of studies examined how estimation frameworks and measurement models influence XNAV performance.

\textcolor{black}{Several works investigated how different sources of uncertainty affect navigation accuracy. Liu et al. (2012) \cite{liu2012pulsar} studied the influence of onboard clock drift and pulsar direction error, modelling these terms as additional states in an augmented unscented Kalman filter and reporting clear improvements. Zhang et al. (2019) \cite{zhang2019orbit} took a related augmented-state approach by combining pulsar timing data with the pulsar position vector, achieving roughly a 75\% accuracy gain over a timing-only XNAV. Xue et al. (2021) \cite{xue2021x} proposed a two-stage estimation method based on Doppler frequency and phase delay and demonstrated faster convergence and smaller errors than a traditional timing-only filter, although the measurement noise was approximated by a single CRLB-derived value.} Similar reliance on CRLB-based assumptions also appears in Chen et al. (2020) \cite{chen2020aspects}. They incorporated pulsar timing noise and
thrusting effects into a particle filter, achieving position error bounds below 30~km, but their analysis relied primarily on the CRLB theoretical bounds and simplified noise models.

Other studies have explored different architectural or sensing choices. Chen et al. (2017) \cite{chen2017autonomous} developed a multirate filtering scheme that processed pulsar measurements with improved efficiency, achieving a root-sum-square position error of 7.5 km after 20 hours of continuous and unobstructed observation of four millisecond pulsars. As part of the PODIUM project, Cacciatore et al. (2022) \cite{cacciatore2023podium} assessed a compact autonomous XNAV unit and found that accuracy below 10 km was feasible when bright pulsars were available. However, their performance analysis primarily considered only detector area, observation duration, pulsed flux, and signal availability, while other critical factors were not accounted for. \textcolor{black}{Zoccarato et al. (2023) \cite{zoccarato2023deep} examined the use of optical pulsars and reached a position accuracy of about 5 km after a five-day simulation, but their study relied solely on the Crab pulsar, which restricts broader applicability.}

\textcolor{black}{Beyond single-sensor designs, hybrid configurations combining pulsar measurements with inertial, optical, or radiometric data have been investigated to improve robustness and overall navigation accuracy \cite{ning2017differential,jiao2016augmentation,xiong2016performance,gu2019optical,zhang2022enhanced}.} Chen and Topputo (2024) \cite{chen2024autonomous} developed a neural-network-assisted XNAV filter in which a radial basis function network predicts the spacecraft state under high-fidelity dynamics, achieving position errors below 3 km, though their work again depended exclusively on the Crab pulsar.

\textcolor{black}{Several researchers have also studied the selection criteria and noise effects when selecting pulsars as navigation sources. In Sheikh’s early work \cite{sheikh2006spacecraft}, timing model imperfections were acknowledged but not included in the navigation system, as the pulsar models were assumed to be sufficiently accurate when estimating position errors. Parthasarathy et al. (2019) \cite{parthasarathy2019timing} analysed glitches and stochastic noise effects on pulsar timing models, and showed that young pulsars often exhibit strong timing noise and frequent glitches, making long-term phase-connected timing solutions difficult to maintain. Mitchell et al. (2018) \cite{mitchell2018sextant}, in the context of SEXTANT, adopted a minimum range uncertainty of about 100 km over a 5000 s observation as a baseline requirement for selecting usable X-ray pulsars. Ray et al. (2017) \cite{ray2017characterization} studied factors that influence TOA measurement errors, compared simulated uncertainties against CRLB estimates, and highlighted the roles of source count rate, background level and pulse shape in determining range measurement uncertainty. They also analysed timing model extrapolation errors and found that the Crab’s ephemeris requires updates approximately every few days. However, similar to \cite{mitchell2018sextant}, their final catalogue selection was driven mainly by range uncertainty, while other criteria that also affect pulsar suitability were not explored in depth. Overall, most of these previous studies have focused primarily on pulsar timing stability, while more detailed and systematic analyses of additional pulsar selection criteria are generally still lacking.}

Despite these advancements in XNAV research, two important limitations persist in the existing studies. First, the measurement noise of pulsar signals is often estimated analytically using signal-to-noise ratio (SNR) formulations or CRLB-based methods. In practice, however, the noise levels associated with real observation data are likely to deviate from these theoretical predictions due to unmodelled instrumental and environmental effects. Second, \textcolor{black}{although previous studies have discussed certain aspects of pulsar selection, these criteria were not examined in a systematic way, and many analyses continued to rely primarily on the Crab pulsar because of its exceptional brightness.} While flux is a key factor, it should not be the sole primary criterion for pulsar selection. Other factors, such as timing stability, sky distribution, pulse profile shape, and visibility, can all significantly affect overall navigation performance. Detailed and systematic analyses of these additional factors remain largely absent in the current literature, highlighting a clear gap that this work aims to address.

Building directly on the identified gaps, this paper adopts an observation-based approach rather than relying on the analytical methods (e.g., SNR models or Cramér-Rao bounds) commonly used in previous studies to estimate measurement uncertainty. Specifically, data collected from the NICER mission are re-scaled to reflect the expected performance of a smaller, customised instrument. In addition, a detailed analysis is conducted on key factors influencing pulsar selection beyond pulsed flux, including pulsar geometric configuration, visibility imposed by solar constraints and Earth shadow, and long-term timing stability. To the best of the authors' knowledge, these aspects have not been systematically addressed in previous literature. As such, this study makes a meaningful contribution towards the development of practical and robust autonomous XNAV systems.



\section{Pulsars} \label{sec: pulsars}


As a pulsar rotates, highly periodic X-ray pulsations are detected from high-temperature regions on its surface (having temperatures peaking in the X-ray band) and/or from the magnetic poles of its plasma-filled magnetosphere. The remarkably stable and predictable evolution of a pulsar's rotation enables its use as an exceptionally accurate celestial clock across a range of applications. While pulsars' rotation slow down over time, the rate of change in the spin period can be measured to high precision over the whole electromagnetic spectrum. More detail on how their rotational stability can be exploited for navigation purposes will be discussed in Section~\ref{sec:timing}. 

Unfortunately however, not all pulsars are suitable for spacecraft navigation. The primary criteria for selecting viable targets for an X-ray pulsar-based navigation system are: (1) the pulsed flux in the X-ray band, which determines the precision with which pulse times-of-arrival (TOAs) can be measured, and (2) the timing stability of the pulsar, that is, how accurately and precisely its rotational phase can be predicted.

Young and energetic pulsars, such as the Crab pulsar, are bright X-ray emitters. Their high flux levels across the X-ray band and narrow pulse profiles enable the measurement of high-quality TOAs over short integration times on relatively small devices, making them attractive candidates for navigation purposes. For instance, a statistical precision of a few tens of microseconds can be achieved for the Crab pulsar from a 5-minute observation with an instrument possessing the capabilities of NICER. However, a significant limitation is their rotational instability and timing noise, particularly pronounced in the Crab pulsar, which restricts the accuracy of extrapolated timing models to only several days. Timing noise refers to stochastic variations in a pulsar's rotational behaviour, typically observed over timescales of days to years. These variations deviate from the otherwise smooth rate of spin-down expected due to the gradual loss of the pulsar's rotational kinetic energy. Timing noise is most commonly observed in young (ages $\lesssim10^4$ years) and energetic pulsars, such as the Crab pulsar, and is generally attributed to changes in the pulsar magnetosphere or complex internal dynamics.

Millisecond pulsars (MSPs), on the other hand, are the ultimate celestial clocks. MSPs are a class of old, rapidly rotating neutron stars (with spin periods $P\lesssim$ 30\,ms) that have been spun up through mass accretion from a binary companion. Due to their high angular momentum and low magnetic fields, MSPs exhibit remarkably stable and predictable rotational behaviour. Unlike young and energetic pulsars, MSPs do not exhibit much stochastic timing noise. Although the majority of MSPs reside in binary systems with another star or compact object, the observed change in spin period caused by their orbital motion can easily be modelled and incorporated into timing models. Consequently, Doppler shifts in pulse arrival times caused by the pulsar's orbit are corrected for when using the timing model to predict TOAs in a given reference frame. Thus, a pulsar being in a binary system does not pose a complication for navigation applications.

Over the past 15 years, extensive timing campaigns of a large sample of MSPs has been carried out to detect low-frequency gravitational waves by searching for correlated timing variations in pulsar data induced by spacetime distortions. This effort, known as the Pulsar Timing Array (PTA) \cite{Detweiler1979} \cite{Hellings1983} \cite{Foster1990} \cite{Hobbs2010}, has produced the most accurate and precise pulsar timing models to date, making them particularly well-suited for spacecraft navigation. However, only a small subset of these MSPs exhibit X-ray pulsations that are sufficiently bright to be detected with compact, lightweight detectors suitable for an XNAV unit.

\begin{figure}[!htbp]
  \centering 
  \includegraphics[width=1.0\linewidth]{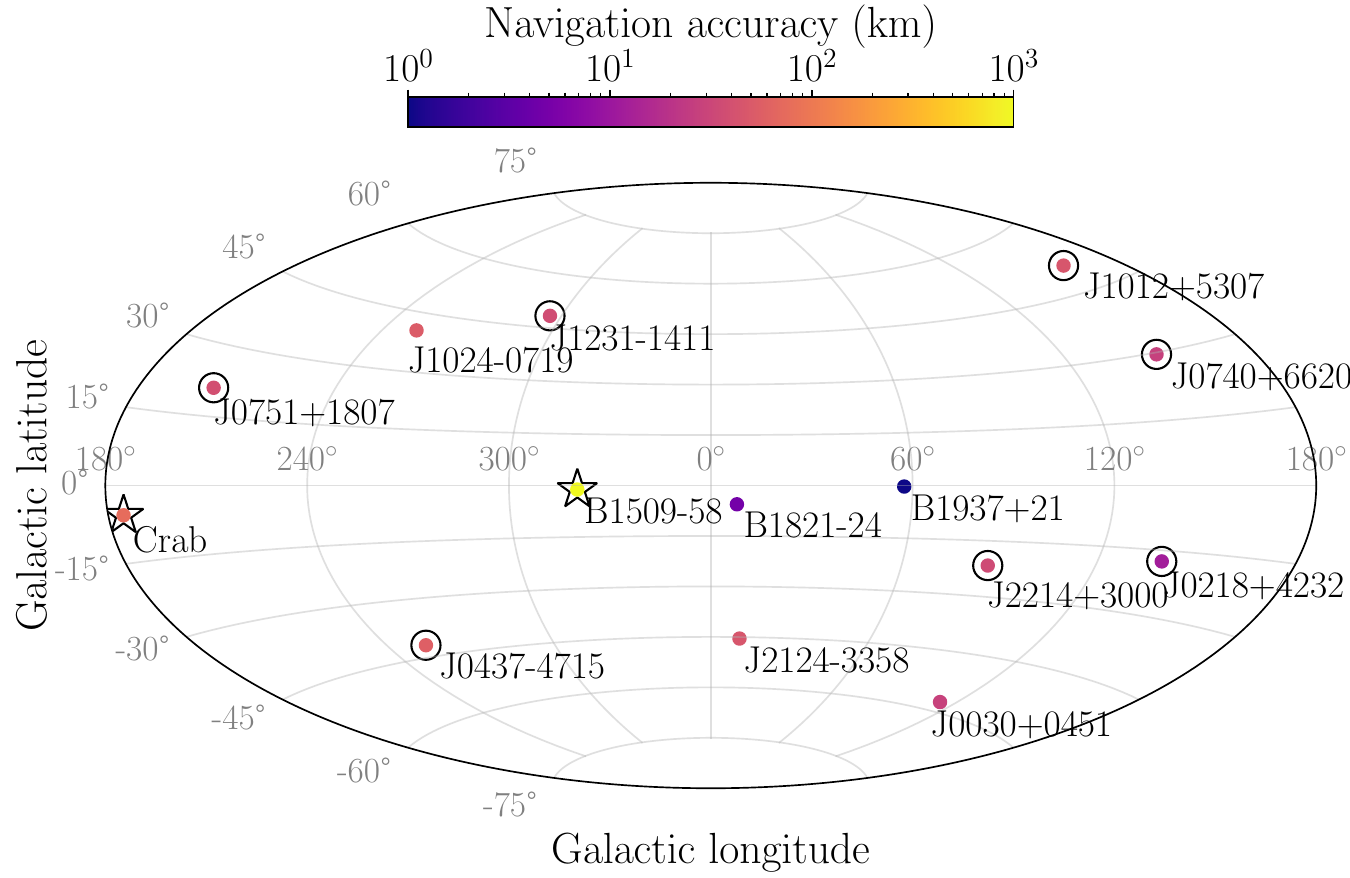}
  \vspace*{-5mm}
  \caption{Spatial distribution of the 14 pulsars included in the catalogue, represented in Galactic coordinates. The two young pulsars, Crab and B1509-58, are indicated by star symbols, while binary pulsars are denoted by black circles. The colour scale represents the maximum achievable navigation accuracy (in km) from a single observation of each pulsar using a detector with an effective area of 200\,cm$\,^2$. For further details, see the observation-based measurement approach described in Section~\ref{sec:meas-observ-approach}.}
  \label{fig:pulsar_map}
\end{figure}

In this work, the same pulsar catalogue utilised by the PODIUM project \cite{cacciatore2023podium} is adopted. This catalogue includes 14 pulsars distributed across the sky, selected for their combination of high rotational stability and strong soft X-ray emission. Their celestial positions, presented in galactic coordinates, are shown in Figure~\ref{fig:pulsar_map}, while Table~\ref{tab:pulsar_parameters} summarises key radiative and timing properties relevant for navigation applications.

\clearpage
\begin{landscape}
\begin{table}
\caption{Summary of key parameters of pulsars.}
\resizebox{1.5\textwidth}{!}{%
\begin{tabular}{cccccccccc}
\toprule
Pulsar Name & Ref. Epoch  & $\nu$ & $\dot{\nu}$ & Right Asc. & Declination & Proper Motion & Orbital Period & $D_0$ & Ref \\
$\mathrm{}$ & $(\mathrm{MJD})$ & $(\mathrm{Hz})$ & $(\mathrm{Hz\,s^{-1}})$ & $(\mathrm{{}^{\circ}})$ & $(\mathrm{{}^{\circ}})$ & $(\mathrm{mas\,yr^{-1}})$ & $(\mathrm{day})$  & $(\mathrm{pc})$ & $\mathrm{}$ \\
\midrule
B0531+21 & 48442.5 & 29.9469230(10) & -3.775350(20) $\times$ 10$^{-10}$ & 5.5755481(14) & 22.014461(17) & 14.8(8) & -- & 1957(309) & \cite{lyne2015Crab} \\
B1509-58 & 55336.0 & 6.59709182778(19) & -6.65310558(27) $\times$ 10$^{-11}$ & 15.2321697(31) & -59.136000(22) & 0.0(0) & -- & 4400(1050) & \cite{parthasarathy2019timing}\cite{verbiest2012pulsar}\\
B1821-24 & 55000.0 & 327.405588060005(6) & -1.7353052(8) $\times$ 10$^{-13}$ & 18.408891087(5) & -24.8696782(15) & 1.3(2.4) & -- & 5368(99) & \cite{reardon2021parkes}\cite{baumgardt2021accurate} \\
B1937+21 & 55599.0 & 641.928232294317(33) & -4.330888(18) $\times$ 10$^{-14}$ & 19.66071146028(17) & 21.583090243(6) & 0.397(6) & --& 3067.5(442.6) &\cite{alam2020nanograv}\cite{reardon2021parkes}\\
J0030+0451 & 55664.0 & 205.530699100590(24) & -4.2976(9) $\times$ 10$^{-16}$ & 0.507618762(8) & 4.86103077(29) & 6.32(10) & -- & 320.5(5.9) & \cite{alam2020nanograv} \\
J0218+4232 & 55000.0 & 430.461061220142(6) & -1.434106(5) $\times$ 10$^{-14}$ & 2.301765918(9) & 42.53816169(16) & 5.93(14) & 2.02884608410(9) & 3150(750) & \cite{du2014very}\cite{verbiest2014distance} \\
J0437-4715 & 55000.0 & 173.68794843058427(6) & -1.72837459(28) $\times$ $10^{-15}$ & 4.62108681478(10) & -47.2525579389(11) & 140.9041(6) & 5.7410448(4) & 156.3(1.3) & \cite{reardon2016timing}\cite{deller2008extremely} \\
J0740+6620 & 57807.0 & 346.53199646083385(34) & -1.463871(12) $\times$ $10^{-15}$ & 7.6793864389(20) & 66.342636823(11) & 32.596(18) & 4.76694461936(8) & 1140(160) & \cite{alam2020nanograv}\cite{fonseca2021refined}\\
J0751+1807 & 55000.0 & 287.4578584521844(31) & -6.4358(4) $\times$ 10$^{-16}$ & 7.852543146(4) & 18.12735691(29) & 13.69(29) & 0.263144266652(4) & 976(207) & \cite{desvignes2016high}\cite{reardon2021parkes}\\
J1012+5307 & 55566.0 & 190.2678373613732(10) & -6.20034(14) $\times$ 10$^{-16}$ & 10.209288326(5) & 53.117294672(26) & 25.685(26) & 0.604672713807(8) & 830(40) & \cite{alam2020nanograv}\cite{ding2020very}\\
J1024-0719 & 56520.0 & 193.7156863607219(6) & -6.96411(8) $\times$ 10$^{-16}$ & 10.4107404240(20) & -7.32212069(4) & 59.68(6) & -- & 1080(40) & \cite{alam2020nanograv}\cite{reardon2021parkes}\cite{kaplan2016psr}\\
J1231-1411 & 55000.0 & 271.453019624388(4) & -1.66705(4) $\times$ 10$^{-15}$& 12.519809194(28) & -14.1954561(8) & 62.34(26) & 1.860143882(9) & 435(45) & \cite{ray2019discovery}\cite{guillemot2016gamma}\cite{bassa2016cool}\\
J2124-3358 & 55000.0 & 202.7938968903887(29) & -8.45958(11) $\times$ 10$^{-16}$ & 21.412179957(5) & -33.97914421(13) & 52.21(8) & -- & 353.9(36.4) & \cite{reardon2021parkes}\\
J2214+3000 & 56920.0 & 320.5922923244547(13) & -1.51382(4) $\times$ 10$^{-15}$ & 22.2441261331(5) & 30.01060953(5) & 20.993(15) & 0.0(0) & 449.7(134.5) & \cite{alam2020nanograv}\\
\bottomrule
\end{tabular}
}
\label{tab:pulsar_parameters}
\end{table}
\end{landscape}
\clearpage

The sample includes two young, bright X-ray pulsars, the Crab pulsar (B0531+21) and B1509-58, which are characterised by high flux levels but comparatively less stable rotation. The remaining 12 objects are nearby MSPs with long-term phase coherence and high-precision timing models. Of these MSPs, seven are members of binary systems. With the exception of J1231-1411, which is not part of the PTA sample, all MSP timing solutions were derived from PTA observations.

\subsection{Pulsar Timing Model} \label{sec:timing}
Pulsar timing is an analytical technique that entails the precise measurement of pulse TOAs through high time-resolution observations. These dedicated observations are typically conducted using radio telescopes and must be performed regularly over multi-year timescales. The measured pulsar signals are subsequently used to derive a deterministic timing model capable of unambiguously predicting the phase of incoming pulses in any reference frame.

A pulsar timing model encodes a set of parameters that can be grouped into two broad categories. The first category comprises the spin parameters, which describe the intrinsic rotation of the neutron star. Specifically, the rotational evolution is modelled as a continuous pulse count function $N(t)$ expressed by the Taylor expansion
\begin{equation}
    N(t) = \sum_{n\geq 1} \frac{\nu^{(n-1)}}{n!}\,(t - t_0)^n + N(t_0)
\label{eq:pulsar_timing_model}
\end{equation}
where $N(t)$ denotes the total (integer) number of rotations elapsed since the reference epoch $t_0$, $\nu \equiv 1/P$ is the rotational frequency, and $\nu^{(n)}$  are the higher-order time derivatives of the rotational frequency. The fiducial pulse count $N(t_0)$ is referenced to a fixed inertial frame, typically the Solar System barycentre (SSB).  The corresponding rotational phase $\phi(t)$ defined on the interval $[0,1)$ is given by the fractional part of $N(t)$

\begin{equation}
    \phi(t) = N(t) - \lfloor N(t) \rfloor
\label{eq:pulsar_timing_model_fractional}    
\end{equation}

The second category of parameters describes the propagation effects that affect the arrival time of pulses as they travel from the pulsar’s comoving frame to an inertial reference frame. For example, the sky position and proper motion of the pulsar must be modelled to account for geometric light travel time across the baseline between the pulsar and the observatory/SSB. Timing models also incorporate corrections for signal propagation through dispersive media, general relativistic frame transformations, and other non-deterministic or stochastic processes, including ionospheric, Solar System, and interstellar plasma dispersion. In cases where the pulsar resides in a binary system, additional orbital parameters are required to model classical and, in some instances, relativistic time delays introduced by the binary motion of the pulsar. For a comprehensive description of these timing model components and their implementation, the reader is referred to \cite{Hobbs2006,ehm06}.  

To summarise, a pulsar timing model is a detailed and rigorously calibrated parameter set that enables transformation of observatory- or spacecraft-based pulse measurements into an inertial reference frame, accounting for the complex dynamics and precision requirements of pulsar timing. Fortunately, much of those details have been incorporated in software packages developed specifically for timing purposes (e.g. \texttt{TEMPO} \cite{tempo}, \texttt{TEMPO2} \cite{tempo2} and \texttt{PINT} \cite{luo2021pint}) that have been thoroughly tested and validated such that systematic errors are negligible. In this work, such software is used to predict the phase of pulse profiles at the SSB at the epoch(s) corresponding to the simulated pulse profile observed by the X-ray detector onboard the navigation unit (Section~\ref{sec:pulsar-profiles}). 

\subsection{Pulsar Pulse Profiles} \label{sec:pulsar-profiles}
Pulsar signals are analysed through their pulse profiles. These are generated by recording the arrival times of individual X-ray photons emitted by a pulsar and detected at the instrument onboard the spacecraft using an atomic clock. Upon striking the detector array, the photon’s energy and time of arrival are measured and recorded. However, the detection of a single photon is insufficient to resolve the periodic nature of the pulsar signal. Instead, a pulse profile is constructed by accumulating a large number of photon arrival events and coherently aligning their arrival times relative to the known timing model of the pulsar (see Section~\ref{sec:timing}). This process is referred to as ``folding''. 

Each pulsar exhibits an energy-dependent pulse profile with a distinctive morphology that remains stable when folded over many rotational cycles of the star. High SNR analytic templates, known as standard profiles $\mathcal{S}(\phi)$, where $\phi = t/P$, are constructed \textit{a priori} by summing profiles from multiple observations. This high-fidelity template can then be phase aligned at the SSB at the reference epoch of the timing model.

In practice, pulsar-based navigation measurements are computed by applying barycentric time correction to the X-ray photons collected onboard (see Section~\ref{sec:meas-observ-approach}), folding these photons at the phase predicted by the timing model, and cross-correlating the observed profiles $\mathcal{P}(\phi)$ with the template profile $\mathcal{S}(\phi)$. The observed profile can be modelled as
 \begin{equation}
     \mathcal{P}(\phi) = a\,\mathcal{S}(\phi - \varphi) + \mathcal{N}(\phi) + b
 \end{equation}
where $a$ is a scaling factor, $\varphi$ is the phase offset between the observed and standard profiles, $\mathcal{N}(\phi)$ describes the noise and $b$ is an amplitude offset. To improve robustness and precision against effects introduced by sampling the pulse profile into discrete time bins, the cross-correlation is usually conducted in the Fourier domain using $\chi^2$-minimisation algorithms \cite{taylor1992pulsar} rather than in the time domain. 

This approach permits high-fidelity extrapolation of the pulsar’s pulse arrival epochs over long time baselines, typically months to years, subject to the accuracy of the fitted parameters in the timing model. This capability is critical for applications in autonomous spacecraft navigation, as it enables prediction and extrapolation of pulsar pulse phases with high temporal fidelity.

\subsubsection{NICER Pulsar Data} \label{sec:nicerdata}
To construct pulse profile templates, publicly-available NICER observations of the pulsars were retrieved. When available, pre-processed datasets from source-specific studies in the literature were used. 

For pulsars without such datasets, raw data products with the longest available exposure times were retrieved from the NICERMASTR database hosted on the HEASARC archive \cite{gendreau2016neutron}. Raw data were processed following NICER’s standard procedures: background screening, filtering and calibration were performed using the \texttt{nicerl2} task in the HEASoft software package \cite{heasoft14}. Time intervals within the South Atlantic Anomaly (SAA; a region where particles produces non-cosmic background radiation) were excluded, as well as overshoot and undershoot events caused by the solar leakage of optical photons and charged particle background interacting with the detector. Information from each measurement power unit (MPU) was combined to produce a cleaned event file. 

Time corrections were then applied to transform photon arrival times from the spacecraft frame to the solar system barycentre, using the \texttt{barycorr} tool and NICER’s orbit solution file in conjunction with the pulsar’s timing model. Rotational phases were assigned to each photon using the \texttt{photonphase} tool of the PINT timing software \cite{luo2021pint}. 

\begin{figure*}[tbp!]
\centering
\includegraphics[width=1\linewidth]{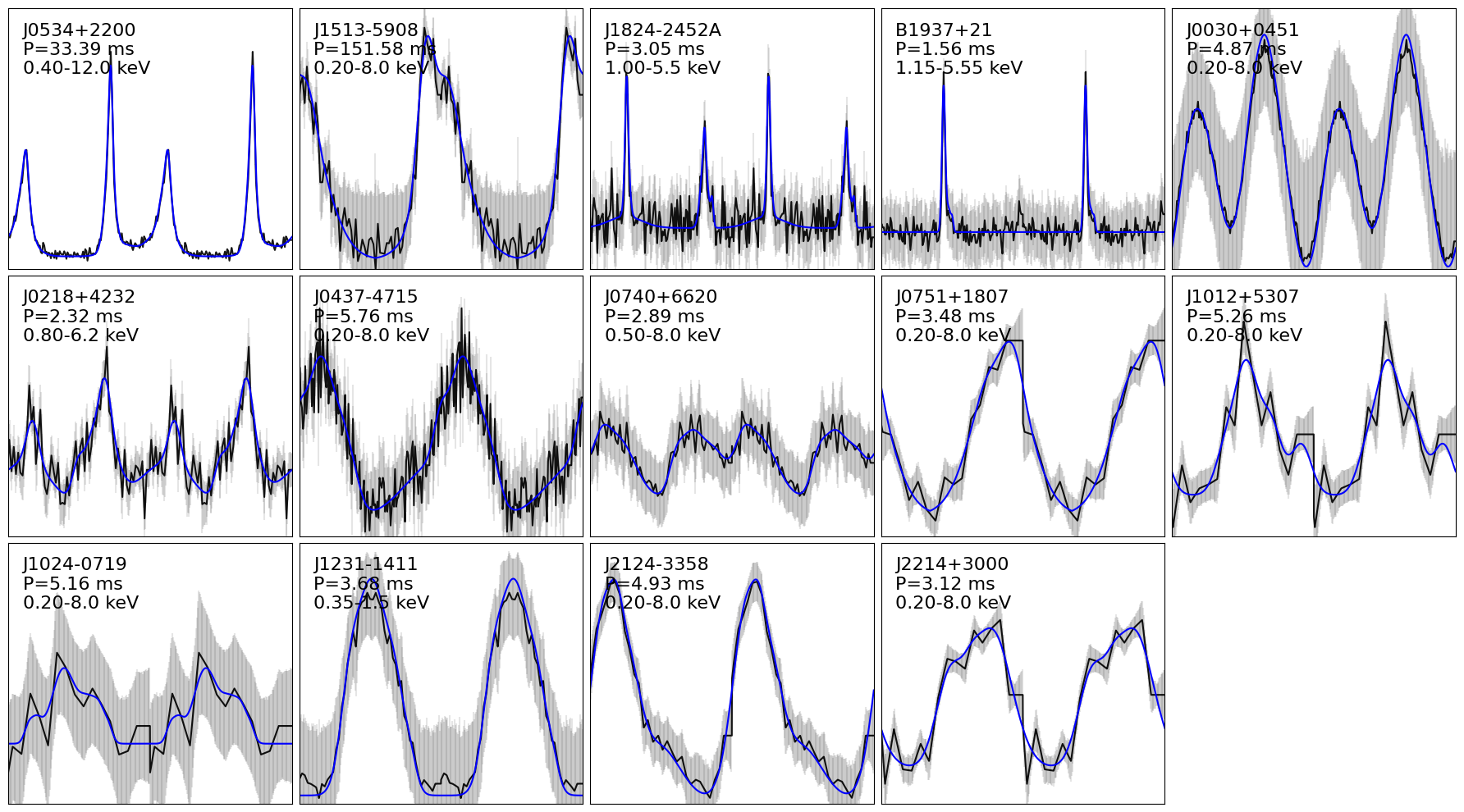}
\caption{Pulse profiles shown over two rotation cycles of the selected 14 pulsars forming the catalogue. The black and shaded grey are the NICER folded profiles and errors, while the blue curves are the Gaussian-composite profile templates. }
\label{fig:template_profile}
\end{figure*}

To enhance the detectability of the pulsed signal, an energy selection was applied based on each pulsar’s spectral characteristics. The optimal energy range was determined by maximising the H-statistic of the resulting pulse profile, thereby improving the signal-to-noise ratio by preferentially selecting photons that contribute most significantly to the periodic modulation.
Time-integrated pulse profiles were then used to construct the final templates. The profiles of all 14 pulsars in the catalogue are shown in Figure~\ref{fig:template_profile}, in which the smooth blue curves are generated by fitting multiple Gaussian functions to observation data from the NICER mission.

\subsection{Pulsar as Navigation Measurement Source}
\label{subsec: pulsar_measurement_source}

Since the pulsar signal is detected as individual photons collected by the X-ray detector onboard a spacecraft, the fundamental measurement of the XNAV system is the photon TOAs. These photon TOAs are subsequently transferred from the spacecraft's location to an inertial reference point, typically the SSB where the pulsar timing model is defined. This time transfer, including higher-order time delay terms, is given by
\begin{equation}
\label{eq:t_transfer}
\begin{split}
    \Delta t (\mathbf{r}) 
    = & \:\: t_{\text{SSB}} - t_{\text{SC}} \\[3pt]
    = & \:\:\frac{\hat{\mathbf{n}} \cdot \mathbf{r}}{c} + \\[3pt]
    & \frac{1}{2cD_0}\left[\left(\hat{\mathbf{n}} \cdot \mathbf{r} \right)^2 - r^2 + 2\left(\hat{\mathbf{n}} \cdot \mathbf{b} \right) \left(\hat{\mathbf{n}} \cdot \mathbf{r} \right) - 2 \left(\mathbf{b} \cdot \mathbf{r} \right) \right] +  \\[5pt] 
    & \frac{2\mu_{\text{Sun}}}{c^3} \ln{\left|\frac{\hat{\mathbf{n}} \cdot \mathbf{r} + r }{\hat{\mathbf{n}} \cdot \mathbf{b} + b} + 1 \right|}
\end{split}
\end{equation}

\noindent where $\hat{\mathbf{n}}$ is the unit direction vector from the SSB to the pulsar, $\mathbf{r}$ is the spacecraft position vector from the SSB, $c$ is the speed of light, $D_0$ is the distance between the pulsar and the SSB, $\mathbf{b}$ is the vector from the Sun centre to the SSB, and $\mu_{\text{Sun}}$ is the gravitational constant of the Sun. 

The first two terms on the right-hand side of Equation~\ref{eq:t_transfer} collectively constitute the Roemer delay, which is a classical geometric delay arising from the difference in light travel time along the line of sight to the pulsar between the spacecraft and the SSB. The third term represents the Shapiro delay, which results from the relativistic curvature of spacetime induced by massive bodies within the Solar System.

\begin{figure}[!htb]
\centering
\includegraphics[width=0.7\linewidth]{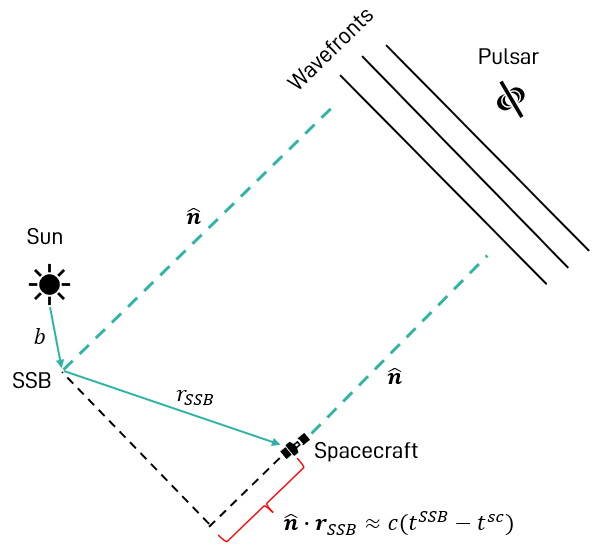}
\caption{Geometrical outline of the XNAV scheme.}
\label{fig:XNAV_scheme}
\end{figure}

If only the first term in Equation~\ref{eq:t_transfer} is retained and the higher-order terms are neglected, the equation is simplified to a geometric representation of the projection of the spacecraft-SSB distance along the pulsar line of sight, as illustrated in Figure~\ref{fig:XNAV_scheme}.

\subsubsection{Signal Processing Pipeline} 

After collecting the photon TOAs as raw measurement data, further processing is required to extract meaningful information that can be incorporated into the navigation filter. The onboard signal processing pipeline is outlined as follows.

\textcolor{black}{Photon arrival times are first recorded by the spacecraft’s X-ray detector. These TOAs are then transferred to the SSB frame using a prior estimate of the spacecraft’s position (through Equation \ref{eq:t_transfer}). At the SSB, the observed profile is generated by folding the TOAs over the pulsar’s period, making use of the timing model (Equations \ref{eq:pulsar_timing_model} and \ref{eq:pulsar_timing_model_fractional}) together with the epoch-folding procedure. A template profile of the same pulsar, constructed from long-term observations, is also available at the SSB. The observed profile is compared with the template to extract the phase shift $\delta\phi$, which reflects the component of the spacecraft’s position error along the pulsar line of sight. This phase shift is then used to refine the initial position estimate. The corresponding correction to the time transfer between the updated spacecraft position and the SSB, based on the prior estimate $\mathbf{r}_{\text{est}}$ and the recovered phase shift $\delta \phi$, is given by}
\begin{equation}
   \Delta t(\mathbf{r}_{\text{updated}}) = \Delta t (\mathbf{r}_{\text{est}}) + \delta \phi \cdot P
\label{eq: phase_shift}
\end{equation}

\begin{figure}
    \centering
    \includegraphics[width=0.8\linewidth]{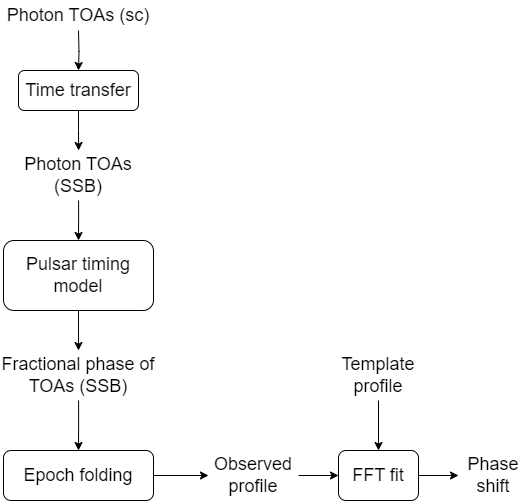}
    \caption{Flowchart of signal processing for XNAV. }
    \label{fig:signal_processing_flowchart}
\end{figure}

\noindent \textcolor{black}{Here, $\Delta t(\mathbf{r}_{\text{est}})$ is computed from the time transfer equation (Equation~\ref{eq:t_transfer}) by substituting the prior position estimate $\mathbf{r}_{\text{est}}$. The overall signal processing sequence described above is summarised in Figure~\ref{fig:signal_processing_flowchart}. }

It is important to note that this method is used to refine only the existing spacecraft position estimate, rather than determining the spacecraft position from scratch, as would be required in a lost-in-space scenario.

\section{Measurement Uncertainty}
\label{sec: meas uncertainty}
\subsection{Analytical Approach}
The measurement accuracy is influenced by several pulsar-specific parameters, including photon flux, background radiation flux, pulsed fraction, pulse width, pulse period etc. In addition, external factors such as instrument size and observation duration also affect the accuracy. Collectively, these factors can be characterised using the signal-to-noise ratio (SNR) of the pulsar signal, as outlined in \cite{sheikh2006spacecraft} and related works:
\begin{equation}
    SNR = \frac{F_x A_x p_f T_{\text{obs}}} {\sqrt{\left[B_x + F_x(1-p_f)\right] \left(A_x T_{\text{obs}} \frac{W}{P}\right) + F_x A_x p_f T_{\text{obs}} }}
\label{eq:SNR}
\end{equation}

\noindent where $F_x$ is the observed X-ray photon flux, $B_x$ is the X-ray background radiation flux, $p_f$ is the pulsed fraction, $W$ is the pulse width, $P$ is the pulse period, $A_x$ is the instrument effective area, and $T_{\text{obs}}$ is the total observation time. The corresponding range accuracy can then be obtained from the timing accuracy using Equation~\ref{eq:range_SNR}, where $c$ denotes the speed of light.
\begin{equation}
    \sigma_{\text{range}} = c \cdot \sigma_{\text{TOA}} = c \cdot \frac{1}{2}\frac{W}{SNR}
\label{eq:range_SNR}
\end{equation}

Alternatively, the measurement uncertainty can also be estimated using the Cramér–Rao Lower Bound (CRLB) method, which provides a theoretical lower bound on the variance of the estimation error. It can be computed using Equation~\ref{eq:CRLB}, as described in \cite{chen2020aspects}.
\begin{equation}
    CRLB = \frac{1}{\lambda_s^2 T_{obs} (1/P)^2} \left[\int_{0}^{1} \frac{s'(\phi)^2}{\lambda_b + \lambda_s s(\phi)} \,d\phi    \right]^{-1}
\label{eq:CRLB}
\end{equation}

\noindent where $\lambda_b$ and $\lambda_s$ denote the background and source photon count rates, respectively. The corresponding range accuracy can then be computed from the CRLB as follows:
\begin{equation}
    \sigma_{\text{range}} = c \cdot \sigma_{\text{TOA}} = c \sqrt{CRLB}
\end{equation}

\begin{figure}[!htb]
\centering
\includegraphics[width=1\linewidth]{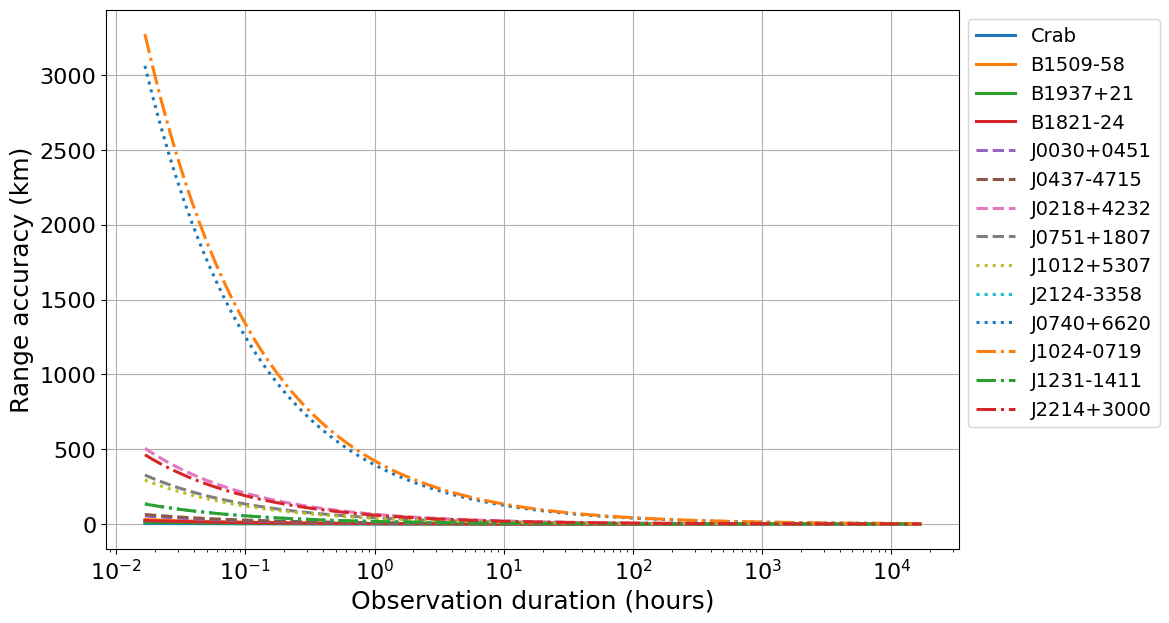}
\caption{Navigation range accuracy as a function of observation duration for an instrument effective area of 200~cm$^2$. Results are obtained using the analytical SNR formulation.}
\label{fig:tobs_range_200cm2_SNR}
\end{figure}

Using the SNR formula in Equations~\ref{eq:SNR} and \ref{eq:range_SNR}, the theoretical range accuracy is plotted as a function of observation duration in Figure~\ref{fig:tobs_range_200cm2_SNR} for all 14 pulsars in the catalogue, assuming a fixed instrument effective area of 200~cm$^2$.

\subsubsection{Effects of $T_{obs}$ and $A_{det}$}

When a specific pulsar is selected as the navigation measurement source, the achievable accuracy depends on both the instrument effective area ($A_{det}$) and the observation duration ($T_{obs}$). Intuitively, increasing either parameter results in a higher number of detected photons. A larger effective area collects more photons per unit time, while a longer observation window allows for more accumulation of photon arrivals. In both cases, the increased photon count improves the SNR of the folded pulse profile, thereby reducing the phase shift uncertainty and ultimately enhancing navigation accuracy.

\begin{figure}[tbp!]
  \centering
  \begin{subfigure}[t]{0.5\textwidth}
    \centering
    \includegraphics[width=1\linewidth,trim={0 0 0 0.04in},clip]{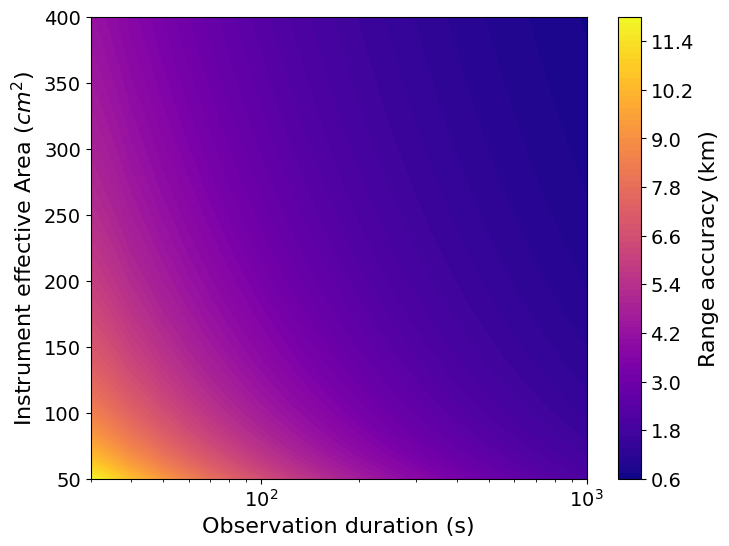}
    \caption{Crab pulsar.}
    \label{figsub:contour_Crab}
  \end{subfigure}
  \begin{subfigure}[t]{0.5\textwidth}
    \centering
    \includegraphics[width=1\linewidth]{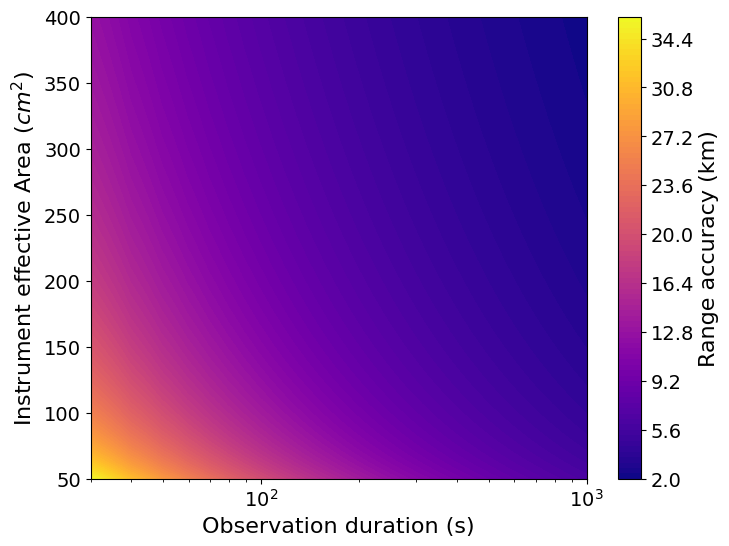}
    \caption{PSR B1937+21.}
    \label{figsub:contour_B1937+21}  
  \end{subfigure}
  \begin{subfigure}[t]{0.5\textwidth}
    \centering
    \includegraphics[width=1\linewidth]{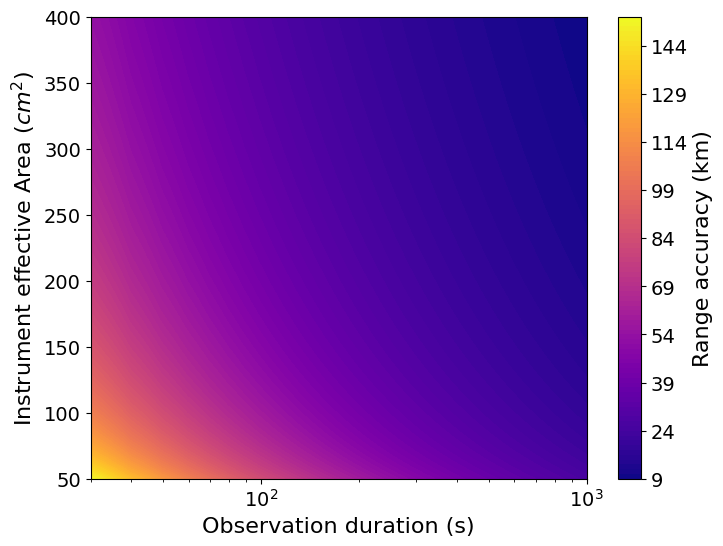}
    \caption{PSR J0030+0451.}
    \label{figsub:contour_J0030+0451}  
  \end{subfigure}
  \caption{Contours of navigation range accuracy as a function of instrument effective area and observation duration, shown for three representative pulsars.}
  \label{fig:contour}
\end{figure}

This relationship is illustrated in Figure~\ref{fig:contour}, which presents simulation results for several pulsars across a range of instrument effective areas and observation durations. For each pulsar, the resulting range accuracy is shown in coloured contour as a function of $A_{det}$ and $T_{obs}$, providing a clear visualisation of how system design choices affect performance. These contour plots serve not only to confirm the expected trends but also to highlight the variation in sensitivity between different pulsar sources.

\begin{figure}[tbp!]
  \centering \includegraphics[width=1\linewidth]{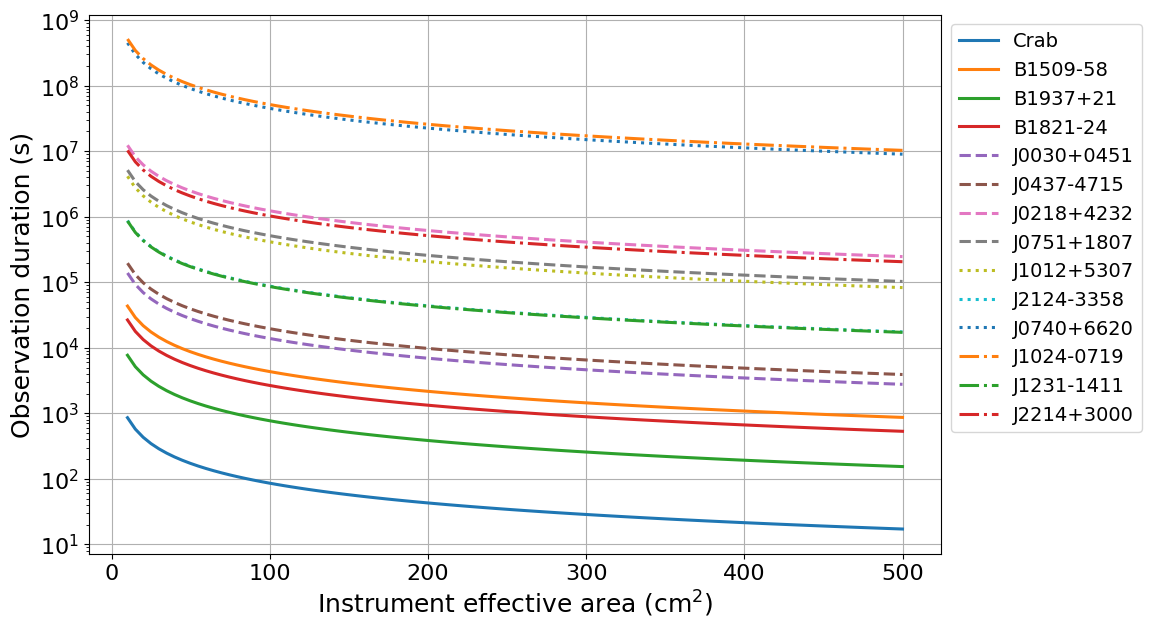}
  \caption{Required observation duration as a function of instrument effective area to achieve 5 km range accuracy, for different pulsars. The Crab pulsar shows the best performance due to its high photon flux.}
  \label{fig:Adet_tobs_5km_sortby_pulsars_SNR}
\end{figure}

To allow a more direct comparison across pulsars, Figure~\ref{fig:Adet_tobs_5km_sortby_pulsars_SNR} shows the required observation duration as a function of instrument effective area, with the range accuracy fixed at 5 km. Each curve corresponds to a different pulsar.  This figure allows for a side-by-side evaluation of the trade-offs associated with different pulsars under the same performance constraint. As shown in the plot, the Crab pulsar consistently outperforms other sources, requiring the shortest observation time and the smallest effective area to meet the target accuracy. This is expected, given its much higher photon flux than other pulsars.

\subsection{Observation-based Approach} \label{sec:meas-observ-approach}

The previous analytical approaches offer a fast and convenient means of obtaining theoretical bounds. However, in the context of an actual space mission, the measurement noise is expected to deviate from these theoretical estimates due to the presence of complex and unpredictable environmental factors. Consequently, it is more informative to derive the measurement uncertainty directly from real observational data, and then adapt the results to match the specifications of the instrument used in this work. This yields a more realistic representation of the measurement uncertainty likely to be encountered in flight.

For this purpose, observational data from the NICER mission is used, which provides X-ray photon count rate, background count rate, and the corresponding energy bands range. These observations also incorporate astrophysical factors such as the galactic hydrogen column density ($N_H$), which affects the photon loss through absorption and scattering along the line of sight. From the NICER data, the observed X-ray energy flux and the background energy flux are computed, both of which can then be converted into photon fluxes. Using this information, the photon TOAs data are simulated for the instrument by applying a non-homogeneous Poisson process. This simulation accounts for the instrument effective area, observation duration, and spacecraft position specific to the mission scenario considered in this work. 

The simulated TOAs are barycentered using the time transfer procedure described in Section~\ref{subsec: pulsar_measurement_source}, and the pulse profile is generated by folding the TOAs at the phase predicted by the pulsar timing model.

Once both the observed and template profiles have been shifted to the phase predicted at the SSB for the same epoch, their absolute alignment is evaluated. Any residual phase offset reflects inaccuracies in the barycentric time transfer corrections (Equation~\ref{eq:t_transfer}). For high-precision phase estimation, the standard technique used in pulsar timing analyses is adopted, which consists of fitting the linear phase gradient in the Fourier domain \cite{taylor1992pulsar}. This method has been implemented in several pulsar timing software packages; in this work, the \texttt{fftfit} tool available in the \texttt{Stingray} Python package is used.

Because each simulated observation is subject to statistical fluctuations from photon counting noise, the uncertainty in the phase offset measurements is estimated using a bootstrap resampling approach. Specifically, for each pulsar and observation duration, 50 independent simulations are performed. The final uncertainty on the phase offset is then quantified as the standard deviation of the resulting distribution of phase measurements. 

In contrast to analytical expressions, which typically assume idealised pulse fractions and widths (as in Equation~\ref{eq:SNR}), this observation-based approach derives these characteristics directly from the full pulse profile generated using NICER data. Consequently, the resulting measurement uncertainty provides a more realistic estimate of accuracy, accounting for both instrumental and astrophysical effects present in an actual mission environment.

\begin{figure}[!htb]
\centering
\includegraphics[width=1\linewidth]{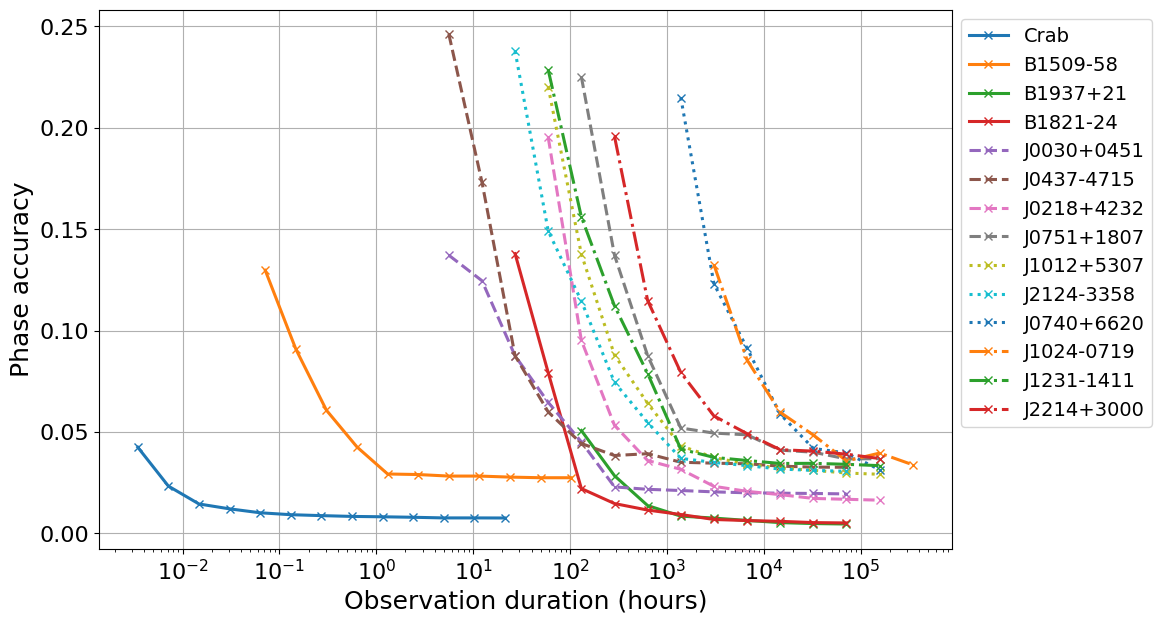}
\caption{Phase accuracy as a function of observation duration for an instrument effective area of 200~cm$^2$. Results are obtained using the NICER data.}
\label{fig:tobs_phase_200cm2_fftfit_14pulsars}
\end{figure}
\begin{figure}[!htb]
\centering
\includegraphics[width=1\linewidth]{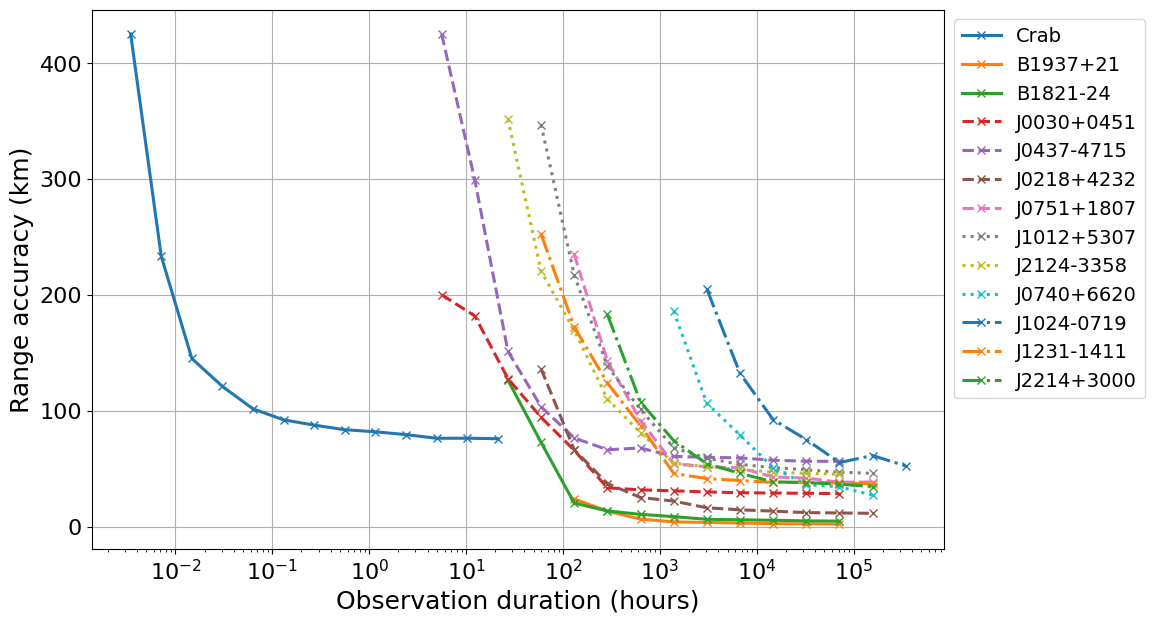}
\caption{Navigation range accuracy as a function of observation duration for an instrument effective area of 200~cm$^2$. Results are obtained using the NICER data.}
\label{fig:tobs_range_200cm2_fftfit_13pulsars}
\end{figure}

Figure~\ref{fig:tobs_phase_200cm2_fftfit_14pulsars} shows the phase shift uncertainty as a function of observation duration, computed using the approach described above. The corresponding range accuracy is obtained by multiplying the phase uncertainty by the pulsar spin period and the speed of light ($\nu_{d} = \nu_{\phi} \cdot P \cdot c$). As a result, even a small phase error $\nu_{\phi}$ can lead to a large range error $\nu_{d}$ for slow-spinning pulsars with large spin periods. This effect is particularly evident for B1509-58 ($P \approx 0.1516$~s) and the Crab pulsar ($P \approx 0.0334$~s), which are the two slowest-spinning sources in the catalogue. For B1509-58, the resulting range error exceeds $10^3$~km, making it unsuitable for high-accuracy navigation. For this reason, it is excluded from the range accuracy results shown in Figure \ref{fig:tobs_range_200cm2_fftfit_13pulsars}, which displays the range accuracy for the remaining 13 pulsars as a function of observation duration.

From Figures~\ref{fig:tobs_phase_200cm2_fftfit_14pulsars} and \ref{fig:tobs_range_200cm2_fftfit_13pulsars}, it is evident that, although the Crab pulsar yields relatively low phase shift uncertainty, its final achievable range accuracy, approximately 65~km, is not among the best. This is primarily due to its comparatively long rotational period, which is roughly an order of magnitude greater than that of other fast-spinning pulsars in the catalogue. As a result, the range error, which scales linearly with the spin period, remains higher despite good phase accuracy.

By comparing the observation-based approach with the analytical approach, Figures~\ref{fig:tobs_range_200cm2_SNR} and \ref{fig:tobs_range_200cm2_fftfit_13pulsars} exhibit consistent trends: the measurement uncertainty decreases as the observation duration increases for a fixed instrument effective area. Both methods indicate that pulsars J1024–0719 and J0740+6620 yield the poorest performance compared to the others. 

However, the results based on the SNR formula suggest that all pulsars can produce measurable range accuracies even with observation durations as short as one minute. This is not realistic. For pulsars with low flux, short observation durations typically result in too few detected photons to form a meaningful observed profile. The resulting data resemble random noise rather than a coherent signal that can be correlated with the template profile to determine phase shifts. This limitation is clearly shown in Figure~\ref{fig:tobs_range_200cm2_fftfit_13pulsars}, where each curve starts at a different observation duration, corresponding to the minimum time required to obtain a meaningful signal for a given instrument effective area of 200~cm$^2$. Since each pulsar has a different flux level, along with variations in pulse profile shape, pulse fraction, and background noise, the minimum observation duration required to achieve a usable range accuracy varies from one pulsar to another. As seen in Figure~\ref{fig:tobs_range_200cm2_fftfit_13pulsars}, pulsars with "better quality" (e.g., higher flux, stronger modulation, lower background noise),  such as the Crab, can achieve measurable range accuracy with shorter observation durations. In contrast, other "lower-quality" pulsars fail to produce reliable results under the same conditions. 

\begin{figure}[!htb]
\centering
\includegraphics[width=1\linewidth]{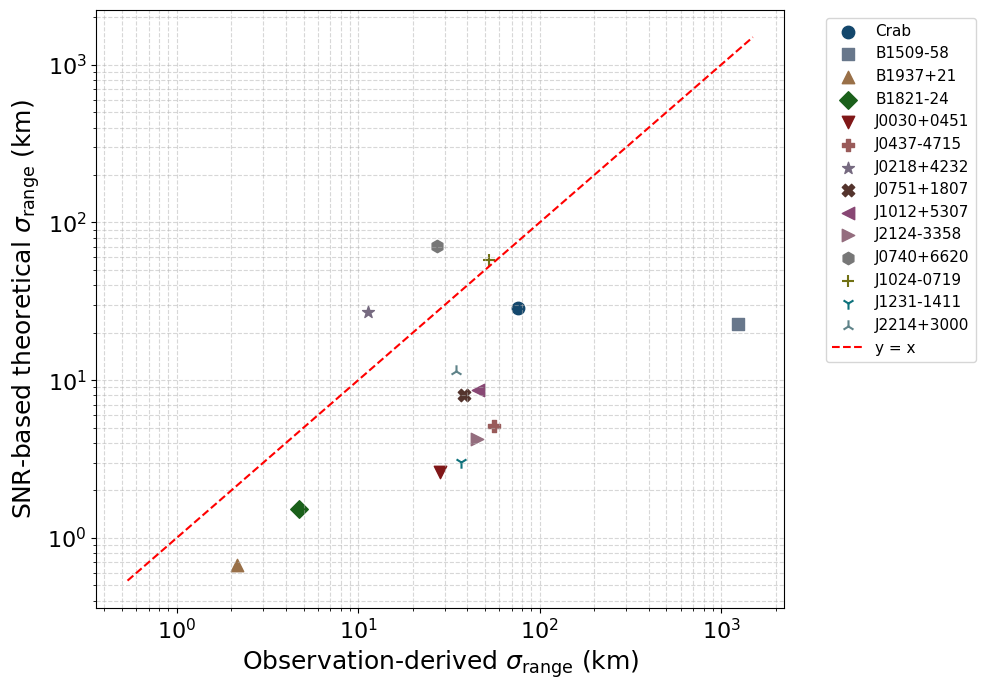}
\caption{Comparison of analytical range accuracy (y-axis) and observation-derived range accuracy (x-axis) for each pulsar. Points lying above the $y=x$ line indicate pulsars where the theoretical estimate over-predicts the observation-derived values, while points below the line indicate under-prediction.}
\label{fig:sigma_range_analytical_obs_compare}
\end{figure}

\textcolor{black}{To allow a more direct comparison between the analytical and observation-derived range accuracies, the final steady values for each pulsar, taken from Figures~\ref{fig:tobs_range_200cm2_SNR} and \ref{fig:tobs_range_200cm2_fftfit_13pulsars} once the curves flatten, are plotted together in Figure~\ref{fig:sigma_range_analytical_obs_compare}. In this plot, points above the $y=x$ line correspond to pulsars for which the analytical estimate is higher than the observation-derived value, while points below the line indicate that the analytical estimate is lower.}

\textcolor{black}{From the distribution of points in Figure~\ref{fig:sigma_range_analytical_obs_compare}, it is clear that the two sets of uncertainties do not match for most pulsars in the catalogue. The differences are also not consistent in direction. A few pulsars, J0218$+$4232, J0740$+$6620 and J1024$-$0719, show analytical uncertainties that are higher than those obtained from the observation-derived results. The remaining pulsars show the opposite trend, where the analytical values fall below the observation-derived uncertainties. Rather than following a clear pattern, the distribution is fairly scattered, and no clear relationship appears that links the analytical uncertainties to the observation-derived ones.}

\textcolor{black}{The result therefore indicates that it is not advisable to use the analytical SNR-based uncertainty and correct it through a simple systematic adjustment, such as adding an offset or applying a scaling factor. The inconsistencies across the pulsar set indicate that the analytical method alone cannot provide a reliable approximation of the measurement uncertainty.}

The discrepancy between the analytical method and the observation-based method highlights the need to account for more realistic instrumental and astrophysical effects when assessing navigation performance. As most existing studies rely solely on analytical methods to estimate measurement uncertainty, the observation-based approach introduced in this paper offers a significant improvement by accounting for instrument-specific and astrophysical effects present in real observations. By incorporating real data from the NICER mission, this method is expected to provide more reliable modelling of onboard performance and enable more realistic performance assessments for future navigation systems.


The Crab pulsar has been chosen as the navigation source in most existing XNAV studies due to its exceptional brightness. However, the Crab's standalone performance does not imply that it will always guarantee optimal navigation results in practical missions. In reality, pulsar selection must account for additional factors such as pulsar direction, visibility constraints, and long-term temporal stability - the latter being a known limitation of the Crab. Moreover, relying on a single pulsar may introduce vulnerabilities due to source occultation or operational constraints. These considerations are addressed in the following section, which explores various criteria for constructing a robust and practical pulsar set for autonomous navigation.

\section{Pulsar Selection Criteria}
\label{sec: pulsar selection factors}

\subsection{Pulsar Geometric Configuration}

To better analyse how navigation accuracy varies along the spacecraft’s trajectory, it is convenient to adopt the radial-transverse-normal (RTN) reference frame, which is centred on and moves with the spacecraft. This local orbital frame provides a physically intuitive basis, in which the radial component (R) points from the central body (e.g., the Sun) towards the spacecraft, the transverse component (T) is aligned with the velocity vector, tangential to the orbit, and the normal component (N) is perpendicular to the orbital plane, completing the right-handed triad.

The transformation from an inertial frame to the RTN frame is constructed using the spacecraft’s inertial position vector $\mathbf{r}$ and velocity vector $\mathbf{v}$, as follows:
\begin{equation}
    R = \frac{\mathbf{r}}{\Vert \mathbf{r} \Vert}, \:\:\:\: N = \frac{\mathbf{r} \times \mathbf{v}}{\Vert\mathbf{r} \times \mathbf{v} \Vert}, \:\:\:\: T = N \times R
\end{equation}

These unit vectors form the rows of the rotation matrix:
\begin{align}
    Q &= \begin{bmatrix}
           R^T \\
           T^T \\
           N^T
         \end{bmatrix},  \qquad \mathbf{x}_{RTN} = Q \cdot \mathbf{x}_{\text{inertial}} 
\label{eq:RTN_rotation}
\end{align}

This transformation allows the position estimation error to be expressed in a frame aligned with the spacecraft’s motion, giving clearer insights into directional variations in navigation performance.

Given their vast distances from the solar system, pulsars can be treated as fixed points on the celestial sphere. Each pulsar is therefore associated with a constant direction vector $\hat{\mathbf{n}}$ in the inertial frame, pointing away from the solar system barycentre. When only a single pulsar is used for navigation, the projection of this fixed direction into the RTN frame changes continuously as the spacecraft moves along its orbit. As a result, the navigation error expressed in RTN components varies over time, depending on the geometry between the spacecraft’s instantaneous motion and the pulsar direction.

\begin{figure}[tbp!]
  \centering \includegraphics[width=1\linewidth]{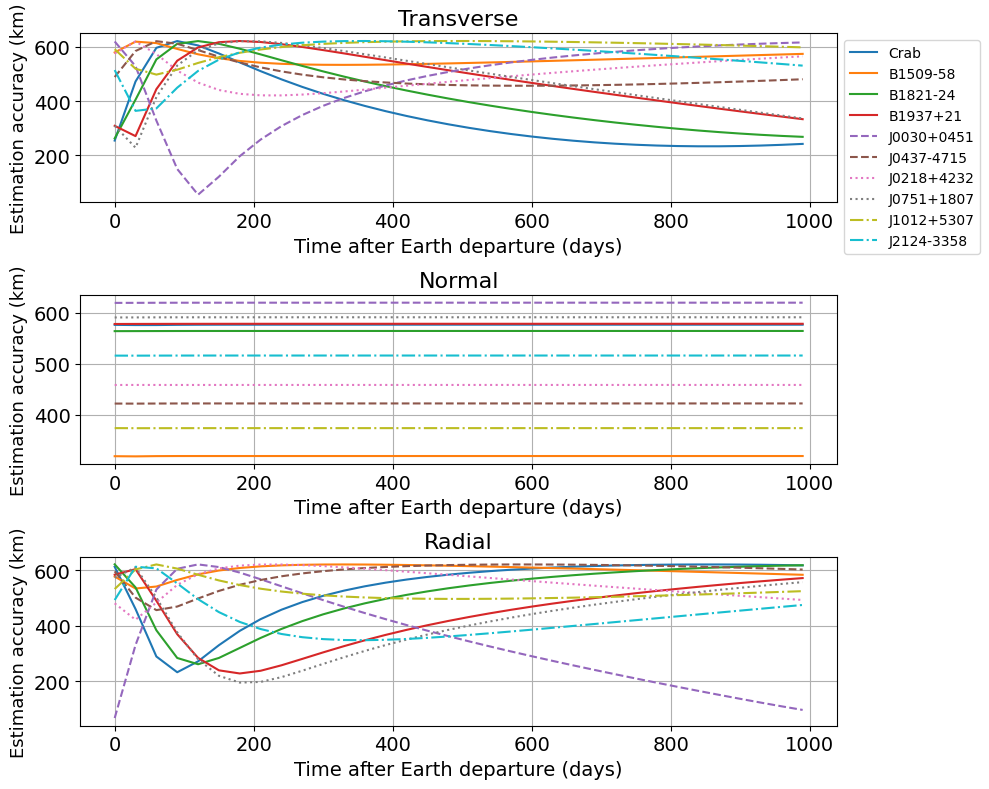}
  \caption{RTN-component position estimation errors along an Earth-Jupiter transfer trajectory when navigating with a single pulsar. The x-axis indicates the spacecraft’s observing position along the trajectory, represented by the time since Earth departure.}
  \label{fig:R-T-N_variation}
\end{figure}

Figure~\ref{fig:R-T-N_variation} illustrates this effect for different pulsars along an Earth-to-Jupiter transfer trajectory. The plot shows how the R-T-N components of the spacecraft's position estimation error vary when observing a pulsar from different locations along the trajectory. The time since Earth departure is used for the x-axis to indicate the spacecraft position along this Earth-Jupiter trajectory.

Interestingly, the normal (N) component remains essentially constant throughout, while the radial (R) and transverse (T) error components change significantly depending on how the pulsar’s direction aligns with the spacecraft’s position. This is because the transfer orbit considered in this case lies close to the ecliptic plane, with minimal out-of-plane motion. As a result, the projection of the pulsar direction onto the normal axis varies very little, leading to a nearly constant contribution to the position error estimation in this direction. In more complex trajectories involving significant inclination changes, the N-component behaviour is expected to differ and should be reassessed accordingly.

\begin{figure}[tbp!]
  \centering \includegraphics[width=1\linewidth]{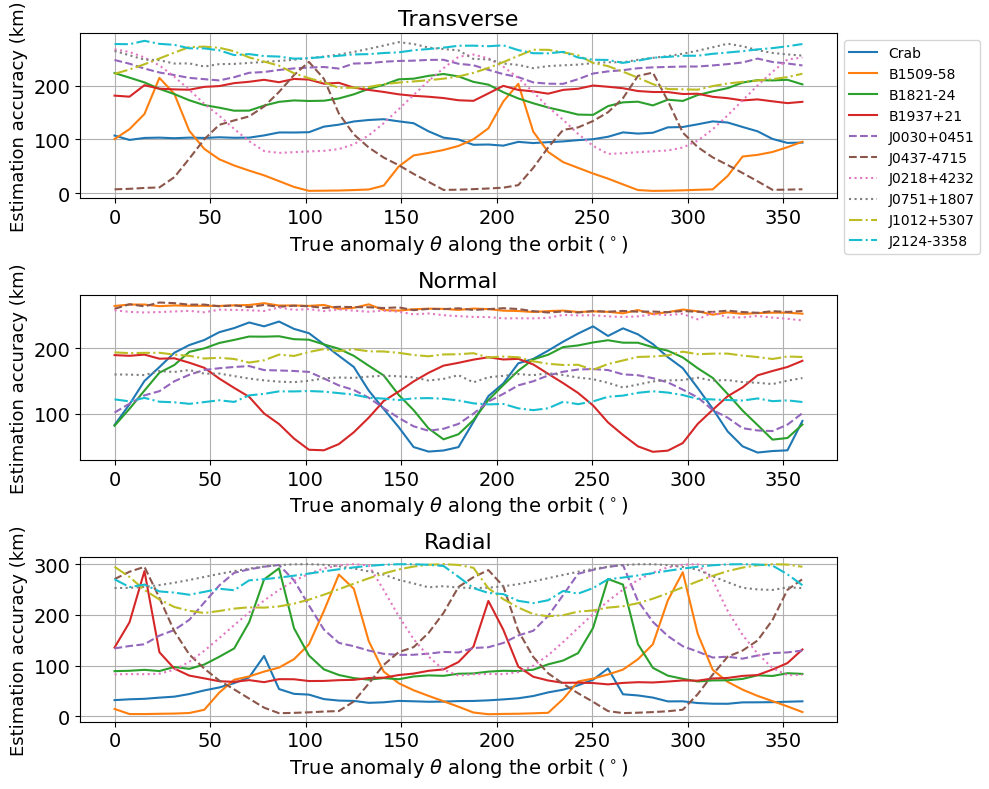}
\caption{RTN-component position estimation errors along a LEO satellite orbit at an altitude of 600~km when navigating with a single pulsar. The x-axis indicates the satellite’s observing position along its orbit, represented by its true anomaly $\theta$.}
  \label{fig:R-T-N_variation_LEO}
\end{figure}

A similar analysis is performed for a low Earth orbit (LEO) satellite in a circular orbit at an altitude of 600~km. Figure~\ref{fig:R-T-N_variation_LEO} shows the variation in position estimation errors along the R, T, and N directions as the satellite observes pulsars from different locations along its orbit. The x-axis represents the satellite's true anomaly, which is a direct indicator of its position along the orbit. It is clear that the position errors in the R, T, and N directions vary significantly depending on the satellite's instantaneous orbital position relative to the observed pulsar's direction.

The results for both the interplanetary and Earth orbits indicate that, for a given pulsar, the resulting navigation error components can vary depending on the observing position. As each pulsar has a distinct direction in space, this suggests that it is possible to select a set of pulsars whose geometric configuration provides a balanced sensitivity across all three spatial directions throughout the mission.

\subsection{Pulsar Visibility}

\subsubsection{Solar Constraint}
Another important consideration in XNAV is the temporary loss of visibility when the Sun obstructs the line of sight to a pulsar. A pulsar becomes temporarily unobservable when its direction falls within a certain angular distance of the Sun from the spacecraft’s perspective. This occurs because the Sun entering the detector’s field of view (FOV) overwhelms the sensor, masking the pulsar signal.

\begin{figure}[tbp]
  \centering \includegraphics[width=0.7\linewidth]{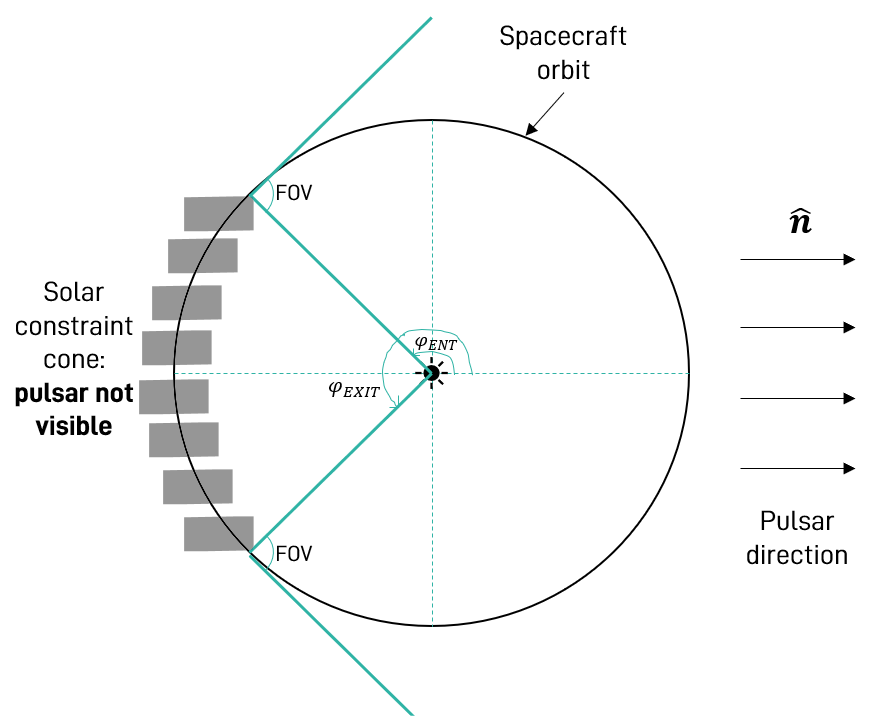}
  \caption{Pulsar visibility under the solar constraint effect in 2D. A pulsar becomes temporarily unobservable when the spacecraft enters the solar constraint cone.}
  \label{fig:solar_constraint_scheme}
\end{figure}

To illustrate this effect, Figure~\ref{fig:solar_constraint_scheme} shows a simplified two-dimensional geometry, where both the Sun and the pulsar lie in the spacecraft’s orbital plane. In this 2D case, visibility is blocked whenever the angular separation between the pulsar and the Sun is smaller than half the detector’s FOV.

\begin{figure}[tbp!]
  \centering \includegraphics[width=0.7\linewidth]{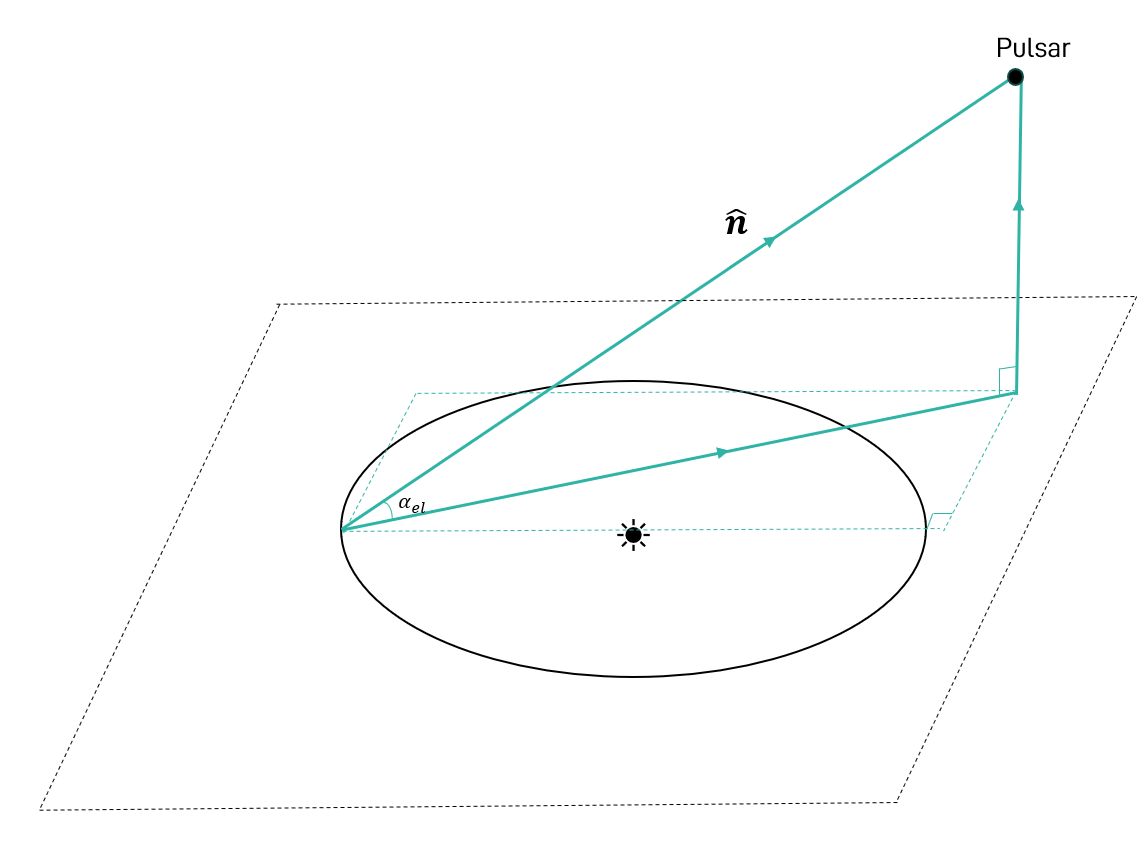}
  \caption{Solar constraint effect in 3D. The pulsar remains continuously visible if it lies sufficiently far above or below the orbital plane.}
  \label{fig:solar_constraint_3D}
\end{figure}

However, the actual constraint is three-dimensional, as pulsars can lie above or below the orbital plane. To capture this geometry, the elevation angle $\alpha_{\text{el}}$ is introduced, defined as the angle between the pulsar’s direction vector $\hat{\mathbf{n}}$ and the spacecraft’s orbital plane. This 3D configuration is illustrated in Figure~\ref{fig:solar_constraint_3D}.

The impact of this constraint depends on the relative geometry between the Sun, the pulsar, and the spacecraft's orbital position. To assess visibility, the pulsar direction vector $\hat{\mathbf{n}}$ is first decomposed into two components: one parallel to the orbital plane and one normal to it. The visibility logic then follows two steps:
\begin{enumerate}
    \item If the elevation angle $\alpha_{\text{el}}$ is greater than half of the detector’s FOV, then the pulsar never enters the Sun-aligned solar constraint cone and is always visible.
    \item If $\alpha_{\text{el}}$ is smaller than half the FOV, the direction vector is projected onto the orbital plane, and a 2D visibility check is performed, as in Figure~\ref{fig:solar_constraint_scheme}.
\end{enumerate}

This approach provides a fast and efficient method to determine visibility interruptions based on known pulsar positions and mission trajectory configuration.

\begin{figure}[tbp]
  \centering \includegraphics[width=1\linewidth]{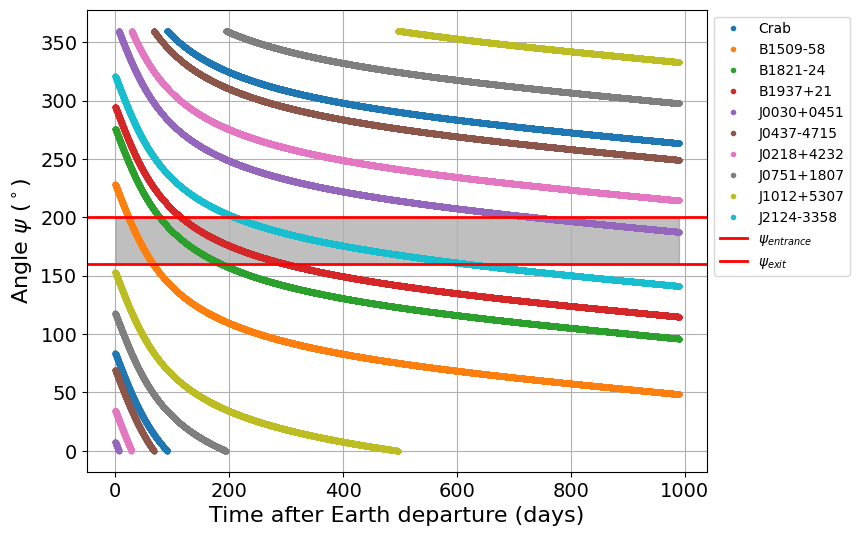}
  \caption{Pulsar visibility during an Earth-Jupiter transfer, limited by the solar constraint (FOV = $40^{\circ}$, start date = 01 Oct 2025). Grey regions indicate times when the pulsar is blocked by the Sun.}
  \label{fig:visibility_solarConstraint_sc}
\end{figure}

To demonstrate how this constraint affects pulsar visibility in practice, Figure~\ref{fig:visibility_solarConstraint_sc} shows visibility windows for several pulsars during an Earth-to-Jupiter transfer mission, assuming a detector FOV of $40^{\circ}$. The time intervals when each pulsar is blocked due to the Sun entering the FOV are shaded in grey.

As shown in the plot, the impact of the solar constraint varies significantly across pulsars. Some remain consistently visible throughout the transfer, while others experience periodic visibility gaps due to alignment with the Sun. These blind spots may compromise navigation performance if not mitigated. In such case, switching to alternate pulsars or blending measurements with other sensors can be considered.

The solar constraint is particularly important during interplanetary missions, where the Sun's apparent motion relative to the spacecraft can cause long-duration outages for certain pulsars. Pulsar selection must therefore balance not only measurement quality (e.g., photon flux) and geometric coverage, but also temporal availability along the trajectory.

\subsubsection{Earth Shadow}
\begin{figure}[tbp]
  \centering \includegraphics[width=1\linewidth]{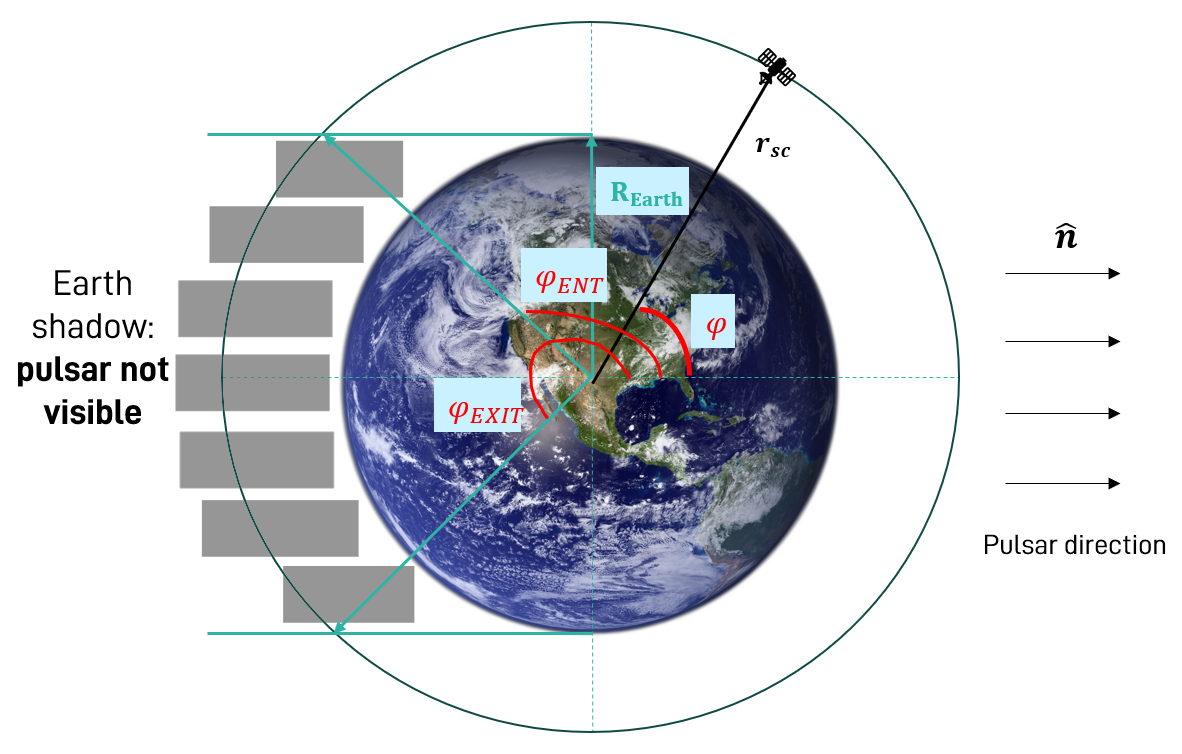}
  \caption{Pulsar visibility under the Earth shadow effect illustrated in 2D.}
  \label{fig:earth_shadow_scheme}
\end{figure}

For a spacecraft in LEO, the Earth's shadow can also have a substantial impact on pulsar visibility. This occurs when the Earth obstructs the line of sight between the spacecraft and a pulsar, typically when the spacecraft is on the far side of the Earth relative to the pulsar. Similar to the solar constraint cone, the Earth shadow is inherently a three-dimensional constraint. However, for simplicity, only the two-dimensional geometry is illustrated in Figure~\ref{fig:earth_shadow_scheme}. The three-dimensional elevation component can be treated analogously to that shown in Figure~\ref{fig:solar_constraint_3D}.

From the figure, the spacecraft is considered to be within the Earth shadow if $\psi_{enter} \leq \psi \leq \psi_{exit}$. This condition can be further expanded using basic geometric relationships \cite{sheikh2006spacecraft}:
\begin{equation}
    \pi - \arccos{\left(\frac{\sqrt{r^2_{sc} - R^2_E}}{r_{sc}} \right)  } \leq \arccos{\left( \hat{\mathbf{n}} \cdot \mathbf{r}_{sc} \right)} \leq \pi + \arccos{\left(\frac{\sqrt{r^2_{sc} - R^2_E}}{r_{sc}} \right)  }
\end{equation} 
\noindent where $R^2_E$ is the average radius of the Earth.



\begin{figure}[tbp]
  \centering \includegraphics[width=1\linewidth]{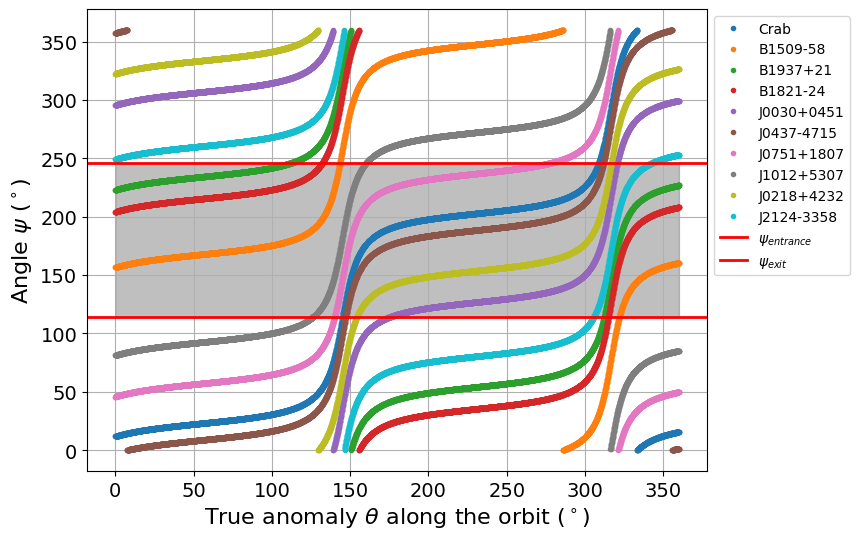}
 \caption{Pulsar visibility due to the Earth shadow for a LEO satellite at an altitude of 600~km (FOV = $40^{\circ}$). Grey regions indicate times when the pulsar is blocked by the Earth.}
\label{fig:visibility_EarthShadow_allpulsars_LEO600km}
\end{figure}

Using this method, the pulsar visibility for a LEO satellite at 600~km altitude is computed and shown in Figure \ref{fig:visibility_EarthShadow_allpulsars_LEO600km}. It is worth noting that although this analysis focuses on Earth orbit, the concept generalises to spacecraft around other planetary bodies, as long as the spacecraft remains sufficiently close to the surface for the shadow geometry to be relevant.

\subsection{Pulsar Timing Stability} \label{sec:psrstability}

The analysis thus far has assumed a perfect pulsar timing model that predicts the pulse phase with no error. In practice, however, the timing model is subject to uncertainties, and phase prediction accuracies tend to degrade over time. This degradation arises both from errors in the deterministic parameters of the model and from stochastic effects such as dispersion measure (DM) variations and intrinsic timing noise.

\begin{figure}[tbp!]
  \centering
  \begin{subfigure}[t]{0.49\textwidth}
    \centering
    \includegraphics[width=1\linewidth,trim={0 0 0 0.04in},clip]{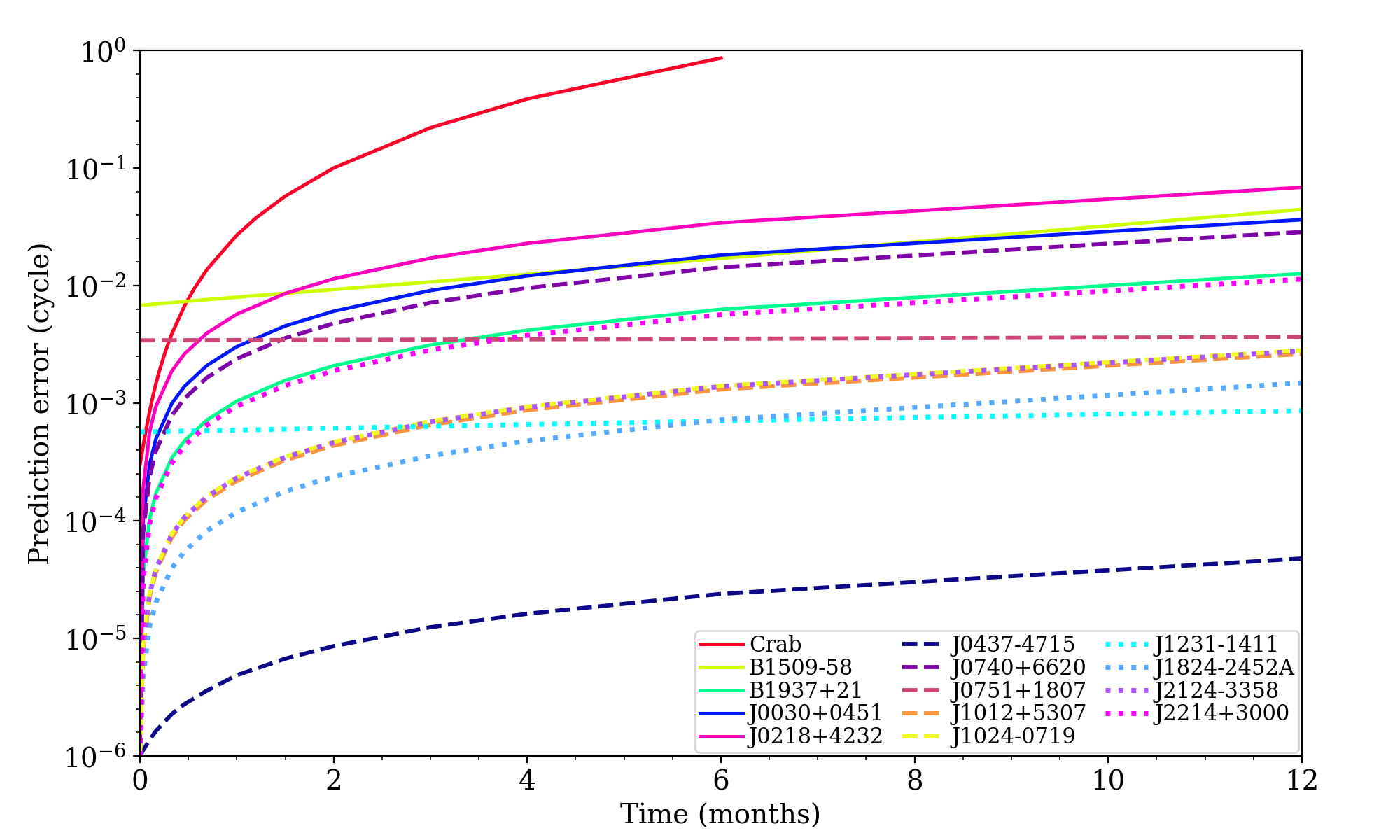}
    \caption{One year extrapolation.}
\label{figsub:extrapolate_model_error_1yr}
  \end{subfigure}
  \begin{subfigure}[t]{0.49\textwidth}
    \centering
\includegraphics[width=1\linewidth]{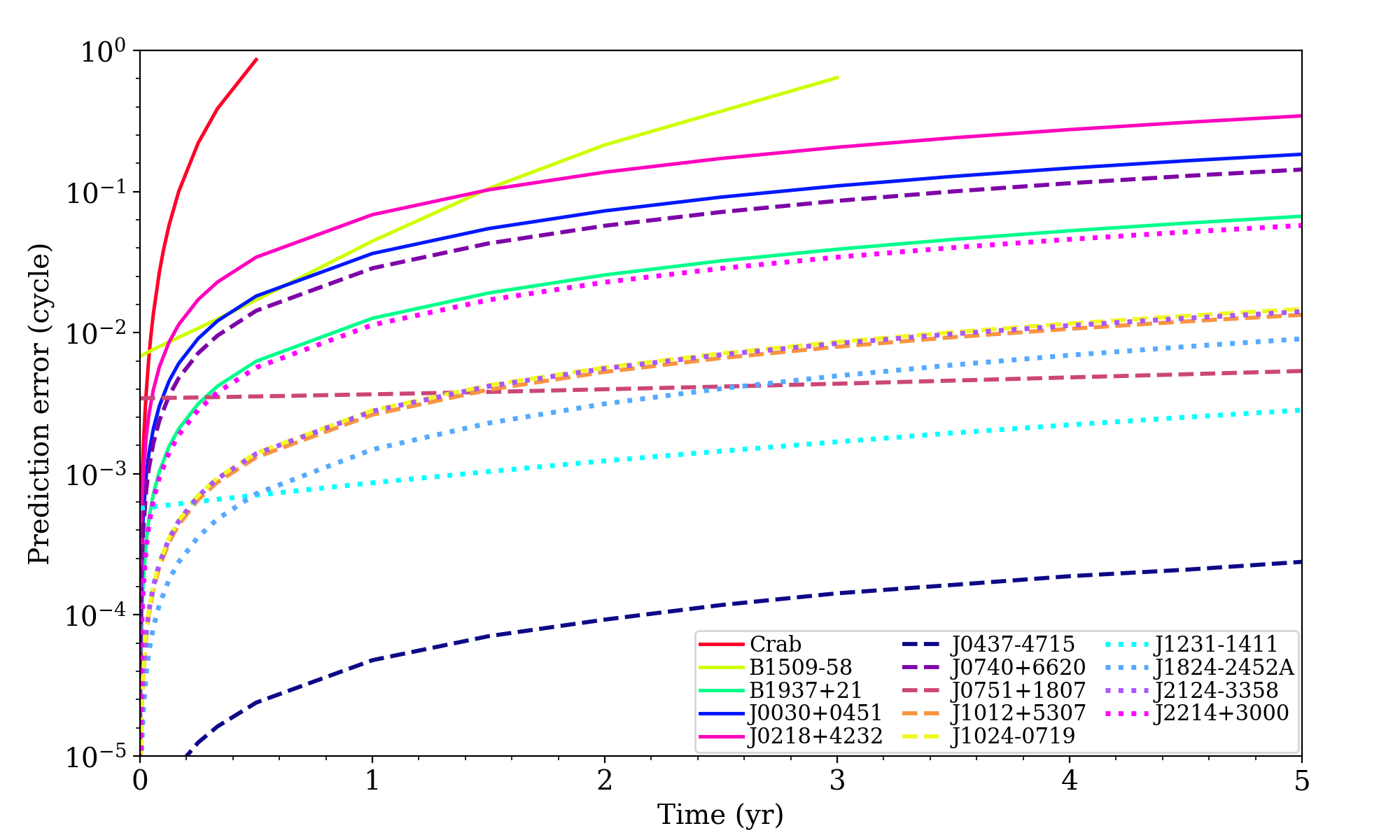}
    \caption{Five years extrapolation.}
\label{figsub:extrapolate_model_error_5yr}  
  \end{subfigure}
  \begin{subfigure}[t]{0.49\textwidth}
    \centering
\includegraphics[width=1\linewidth]{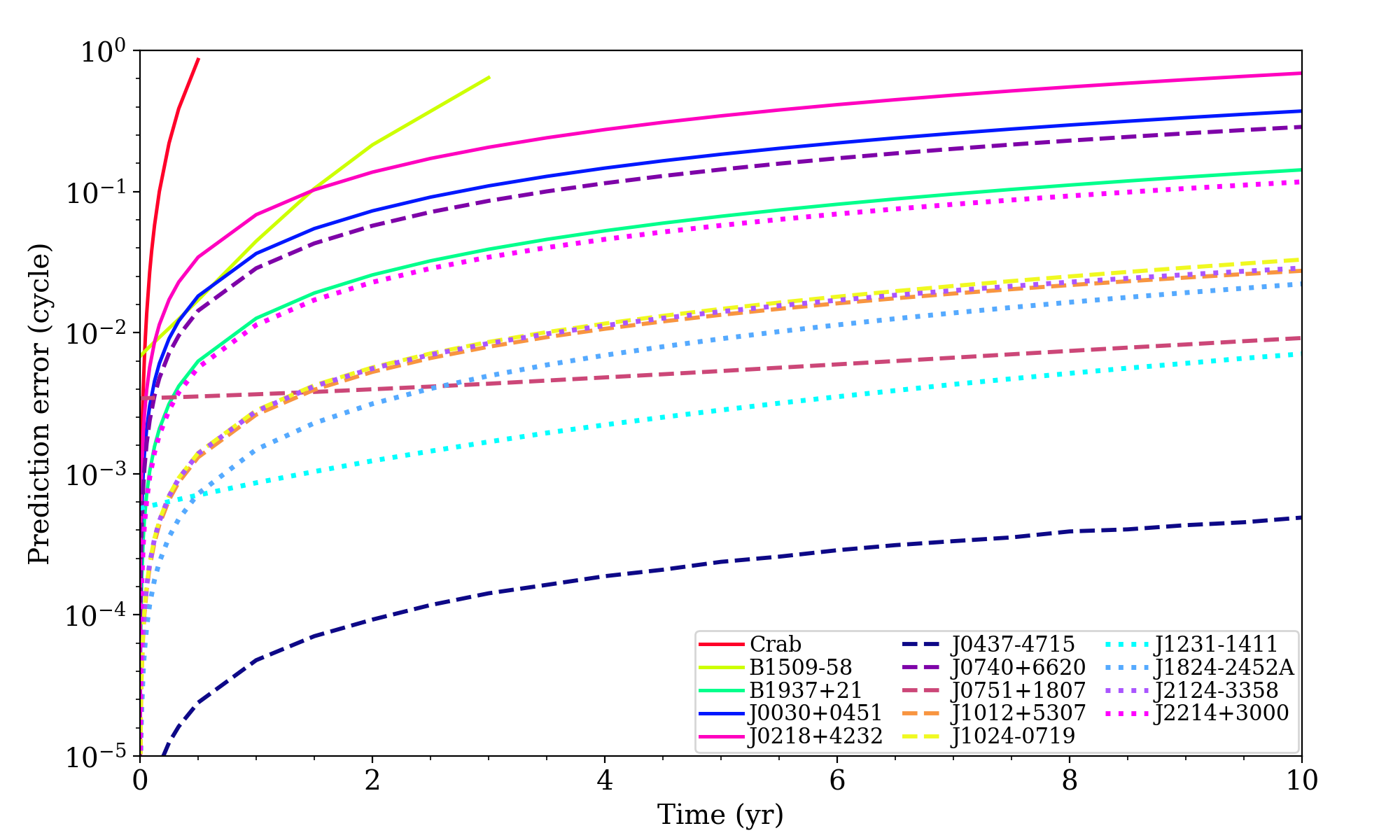}
    \caption{Ten years extrapolation}
\label{figsub:extrapolate_model_error_10yr}  
  \end{subfigure}
   \begin{subfigure}[t]{0.49\textwidth}
    \centering
\includegraphics[width=1\linewidth]{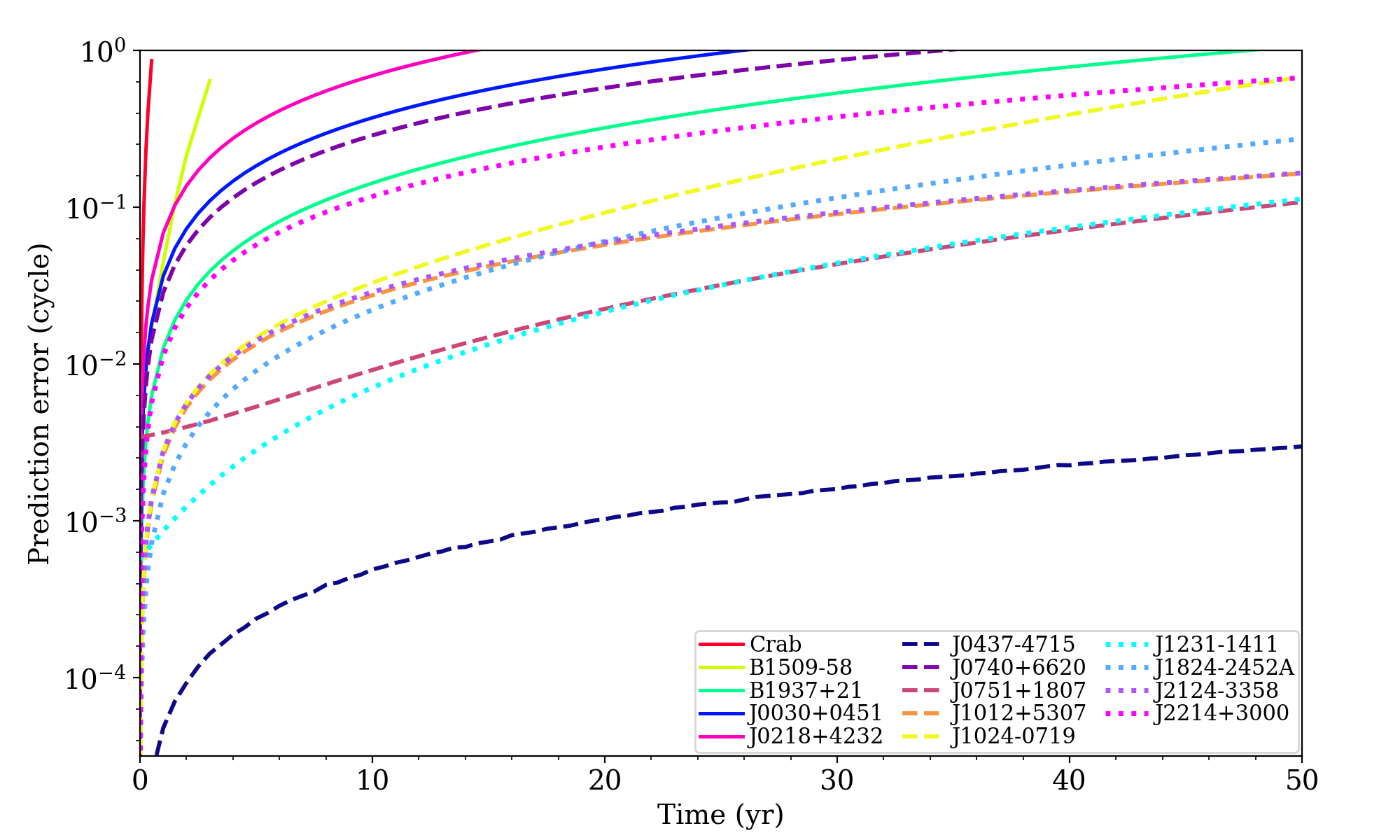}
    \caption{Fifty years extrapolation}
\label{figsub:extrapolate_model_error_50yr}  
  \end{subfigure}
  \caption{Growth of pulsar phase prediction error due to timing model extrapolation.}
  \label{fig:extrapolate_model_error}
\end{figure}

To characterise how prediction accuracy degrades over time, several randomised timing models are generated by randomly selecting the known model parameters within their respective uncertainty bounds. Stochastic DM fluctuations are also included in the process. These models are then extrapolated forward in time, and the pulse phases are estimated at the same reference radio frequency of the true model. The spread in these predicted pulse phases provides an estimate of the total error budget in phase predictions as a function of extrapolation time. The result is illustrated in Figure~\ref{fig:extrapolate_model_error}, which shows how phase prediction errors accumulate when the model is extrapolated forward in time without updates. 

Some pulsars are inherently more stable than others. Older pulsars, especially those that have undergone sustained accretion in a binary system and have been spun up to millisecond periods, tend to exhibit lower levels of timing noise and are therefore more suitable for long-duration applications. In contrast, younger pulsars often show greater variability in their timing behaviour, which limits the length of time for which their models remain reliable without updates.

These phase prediction errors directly impact the navigation solution, since the estimated position of the spacecraft depends on accurately matching the observed pulse phase to the expected one. As a result, the long-term performance of a pulsar-based navigation system is constrained by the stability and accuracy of the timing models used for each pulsar. To ensure that phase prediction errors remain within acceptable bounds, regular updates to the timing models are required. These updates are typically based on Earth-based radio observations and must be communicated to the spacecraft via telemetry. The frequency of such updates will depend on the specific pulsars used, their timing stability, and the level of phase precision required by the navigation system.

\section{X-ray Pulsar Navigation}
\label{sec: XNAV}

With the key pulsar selection criteria defined, these are applied within a spacecraft navigation context to assess the overall performance of the XNAV system. To simulate the pulsar signals, photon arrival times from each pulsar are modelled as a non-homogeneous Poisson process (NHPP), where the rate function is defined by the pulsar’s template pulse profile. This enables the generation of realistic photon TOAs data consistent with the pulsar’s observed behaviour.

To simulate and test the XNAV system, two scenarios are considered. The first involves a spacecraft on an interplanetary transfer from Earth to Jupiter, with a transfer duration of approximately three years. The second involves a LEO satellite in a circular orbit at an altitude of 600~km. The spacecraft properties are assumed to be identical in both cases and are summarised in Table~\ref{tab:sc_properties}.

\begin{table}[tb]
\caption{Spacecraft properties used in the simulation.}
\centering
\begin{tabular}{l|c}
\hline
Mass $m$ (kg)                 & 100 \\
Spacecraft cross-sectional area $A$ (m$^2$) & 5   \\
Instrument effective area $A_{det}$ (cm$^2$) & 200   \\
Drag coefficient $C_D$         & 2.2 \\
Reflectivity coefficient $C_R$ & 1.3 \\ \hline
\end{tabular}
\label{tab:sc_properties}
\end{table}

\subsection{Spacecraft Dynamics}
The state vector used by the navigation filter to describe the spacecraft dynamics comprises its three-dimensional inertial position and velocity:
\begin{align}
    \mathbf{x}(t)  &= \begin{bmatrix}
          \mathbf{r}(t) \\
           \mathbf{v}(t) 
         \end{bmatrix}
\end{align}

In a general form, the governing equation for the state dynamics can be written as:
\begin{equation}
    \dot{\mathbf{x}}(t) = \mathbf{f}(t,\mathbf{x}(t)) + \mathbf{w}(t)
\label{eq:dynamics}
\end{equation}

\noindent where $\mathbf{w}(t)$ is the process noise, and $\mathbf{f}(t,\mathbf{x}(t))$ defines the deterministic part of the spacecraft dynamics. In this case, the function is given by:
\begin{align}
    \mathbf{f}(t,\mathbf{x}(t)) &= \begin{bmatrix}
           \mathbf{v}(t) \\
           \mu\frac{\mathbf{r}}{r^3} + \mathbf{a}_{\text{pert}}
         \end{bmatrix}
\end{align}

Here, $\mathbf{a}_{\text{pert}}$ denotes the sum of perturbation accelerations acting on the spacecraft. In the interplanetary case, the dominant contributions are solar radiation pressure and third-body gravitational effects from other planetary bodies in the solar system. While for the LEO case, the main perturbations are atmospheric drag and the Earth's oblateness, modelled through the $J_2$ term.

\subsection{Measurement Model}
Similar to the state dynamics, the measurements can also be expressed as a function of the system state:
\begin{equation}
    \mathbf{y}(t) = \mathbf{h}(t, \mathbf{x}(t)) + \boldsymbol{\nu}(t)
\end{equation}

\noindent where $\boldsymbol{\nu}(t)$ denotes the measurement noise, and $\mathbf{h}(t, \mathbf{x}(t))$ is the measurement model that relates the observations to the system state $\mathbf{x}(t)$.

As discussed in Section~\ref{subsec: pulsar_measurement_source}, for each pulsar, the post-processed measurement model can be written as the TOA difference between the spacecraft position and the SSB:
\begin{equation}
    \begin{split}
    h(\mathbf{x}) = \: &\Delta t (\mathbf{r}) \\[3pt]
    = & \:\:\frac{\hat{\mathbf{n}} \cdot \mathbf{r}}{c} + \\[3pt]
    & \frac{1}{2cD_0}\left[\left(\hat{\mathbf{n}} \cdot \mathbf{r} \right)^2 - r^2 + 2\left(\hat{\mathbf{n}} \cdot \mathbf{b} \right) \left(\hat{\mathbf{n}} \cdot \mathbf{r} \right) - 2 \left(\mathbf{b} \cdot \mathbf{r} \right) \right] +  \\[5pt] 
    & \frac{2\mu_{\text{Sun}}}{c^3} \ln{\left|\frac{\hat{\mathbf{n}} \cdot \mathbf{r} + r }{\hat{\mathbf{n}} \cdot \mathbf{b} + b} + 1 \right|}
\end{split}
\end{equation}

\subsection{Navigation Filter}
An extended Kalman filter (EKF) is implemented for the spacecraft state estimation. The filter operates in two steps: a time-update step that propagates the state forward, and a measurement-update step that incorporates pulsar observations to correct the state estimate. The spacecraft dynamics is modelled as a continuous-time system, whereas the measurements are available at discrete time intervals. As a result, the process and measurement equations are written as follows:
\begin{equation}
     \dot{\mathbf{x}}(t) = \mathbf{f}(t,\mathbf{x}(t)) + \mathbf{w}(t)
\end{equation}
\begin{equation}
     \mathbf{y}_k = \mathbf{h}(\mathbf{x}_k) + \bm{\nu}_k
\end{equation}

The filter is initialised by specifying $\mathbf{x}_0^{+}$ and $\mathbf{P}_0^{+}$:                   
\begin{equation}
    \mathbf{x}_0^{+} = E[\mathbf{x}_0]
\end{equation}
\begin{equation}
     \mathbf{P}_0^{+} = E\left[ \left(\mathbf{x}_0-\mathbf{x}_0^{+} \right)  \left(\mathbf{x}_0-\mathbf{x}_0^{+} \right)^T \right]
\end{equation}

The EKF time-update (prediction) step is outlined as follows:
\begin{equation}
    \mathbf{x}_k^{-} = \mathbf{x}_{k-1}^{+} + \int_{t_{k-1}}^{t_k} \mathbf{f}(t,\mathbf{x}(t)) \:\: dt
\label{eq:state_prediction}
\end{equation}
\begin{equation}
    \mathbf{P}_k^{-} = \mathbf{P}_{k-1}^{+} + \int_{t_{k-1}}^{t_k} \mathbf{F}\mathbf{P} + \mathbf{P}\mathbf{F}^T + \mathbf{Q} \:\: dt
\end{equation}

At time $t_k$, once the measurement $y_k$ is available, it is incorporated into the EKF measurement-update (correction) step to update both the state and covariance estimates:
\begin{equation}
    \mathbf{K}_k = \mathbf{P}_k^{-} \mathbf{H}_k^T \left(\mathbf{H}_k \mathbf{P}_k^{-} \mathbf{H}_k^T + \mathbf{R}_k \right)^{-1}
\end{equation}
\begin{equation}
    \mathbf{x}_k^{+} = \mathbf{x}_k^{-} + \mathbf{K}_k \left( y_k - h_k  \right)
\label{eq:correction_meas}
\end{equation}
\begin{equation}
    \mathbf{P}_k^{+} = (\mathbf{I} - \mathbf{K}_k\mathbf{H}_k)\mathbf{P}_k^{-} (\mathbf{I} - \mathbf{K}_k\mathbf{H}_k)^T + \mathbf{K}_k \mathbf{R}_k \mathbf{K}_k^T
\end{equation}

In the above expressions, the Jacobian matrices $\mathbf{F}$ and $\mathbf{H}$ are defined as $\mathbf{F}_k = \left.\frac{\partial{\mathbf{f}(\mathbf{x})}}{\partial{\mathbf{x}}}\right\vert_{\mathbf{x} = \mathbf{x}_{k-1}^{+}}$ and $\mathbf{H}_k = \left.\frac{\partial{h(\mathbf{x})}}{\partial{\mathbf{x}}}\right\vert_{\mathbf{x} = \mathbf{x}_k^{-}}$ respectively. $\mathbf{Q}$ is the process noise covariance matrix and $\mathbf{R}$ is the measurement noise covariance matrix.

\subsection{Navigation Performance}
\subsubsection{Case 1: Earth-Jupiter Interplanetary Spacecraft}

\begin{figure}[tbp!]
  \centering \includegraphics[width=0.6\linewidth]{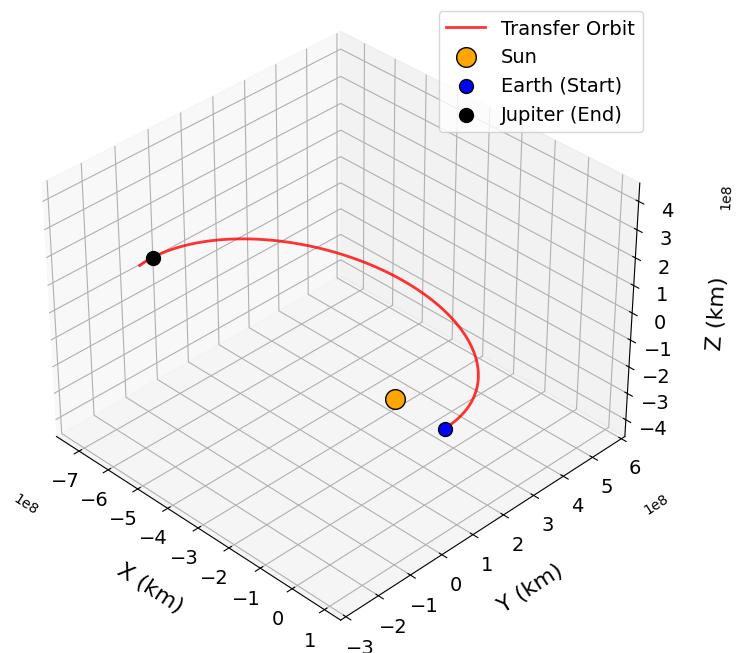}
  \caption{Earth-Jupiter interplanetary transfer orbit with a transfer duration of about three years.}
\label{fig:HomannTransfer_EarthJupiter}
\end{figure}

The navigation simulation is first run for the Earth-Jupiter interplanetary transfer scenario. The spacecraft’s trajectory is shown in Figure~\ref{fig:HomannTransfer_EarthJupiter}, while the launch and arrival dates, along with the initial orbital conditions, are provided in Table~\ref{tab:launch_conditions}. It is assumed that a suitable launch window has been selected to ensure proper planetary alignment for the transfer.

\begin{table}[!]
\caption{Launch and arrival dates for the Earth-Jupiter transfer orbit, along with the spacecraft initial orbital conditions for simulation.}
\centering
\begin{tabular}{cccc}
\toprule
\multicolumn{2}{c}{Launch date}             & \multicolumn{2}{c}{01 Oct 2025} \\
\multicolumn{2}{c}{Arrival date}            & \multicolumn{2}{c}{30 May 2029} \\ 
\midrule
$x_0$ (km) & \multicolumn{1}{c}{1.495979e+08} & $vx_0$ (km/s)     & 0.000000e+00     \\
$y_0$ (km) & \multicolumn{1}{c}{0.000000e+00} & $vy_0$ (km/s)     & 3.857571e+01     \\
$z_0$ (km) & \multicolumn{1}{c}{0.000000e+00} & $vz_0$ (km/s)     & 0.000000e+00     \\ 
\bottomrule
\end{tabular}
\label{tab:launch_conditions}
\end{table}

To assess the performance of the proposed navigation system, a segment of this transfer orbit during the first 365 days is considered.  For simplicity, only pulsars that remain continuously visible throughout this time interval are considered. Based on the visibility analysis shown in Figure~\ref{fig:visibility_solarConstraint_sc}, the following pulsars are excluded due to solar constraints: B1509-58, B1821-24, B1937+21, and J2124-3358.

From the remaining set, pulsars that possess both high photon flux and favourable directional sensitivity are prioritised, with the aim to ensure balanced navigation accuracy across the radial, transverse, and normal directions in the spacecraft’s RTN body frame. In addition, pulsars that have relatively stable timing behaviour over the one-year interval are considered. Based on these criteria, the following three potential sets of pulsars are selected and the corresponding navigation systems' performance is tested and compared:
\begin{itemize}
    \item Combination \#1: Crab, J0030+0451, J0437-4715
    \item Combination \#2: J2124-3358, J0030+0451, J0437-4715
    \item Combination \#3: J1012+5307, J0030+0451, J0437-4715
\end{itemize}

\begin{figure}[tbp!]
  \centering
  \begin{subfigure}[t]{0.49\textwidth}
    \centering
    \includegraphics[width=1\linewidth,trim={0 0 0 0.04in},clip]{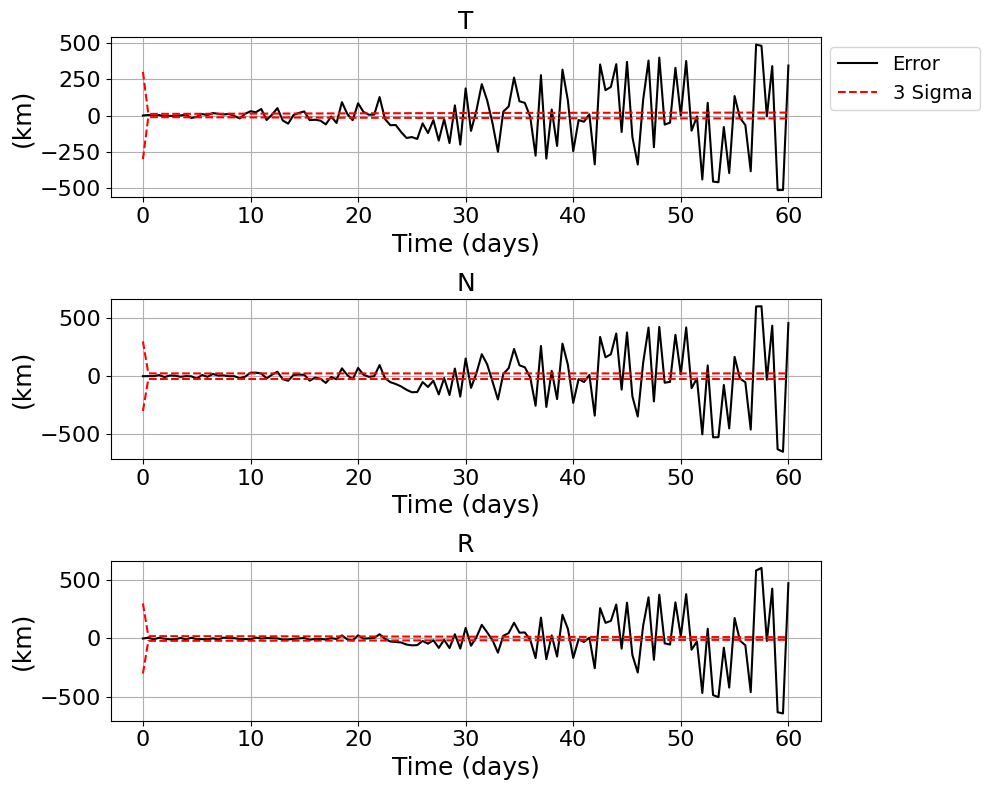}
    \caption{Position estimation error.}
\label{figsub:filter_pos_crab_J0030_J0437}
  \end{subfigure}
  \begin{subfigure}[t]{0.49\textwidth}
    \centering
\includegraphics[width=1\linewidth]{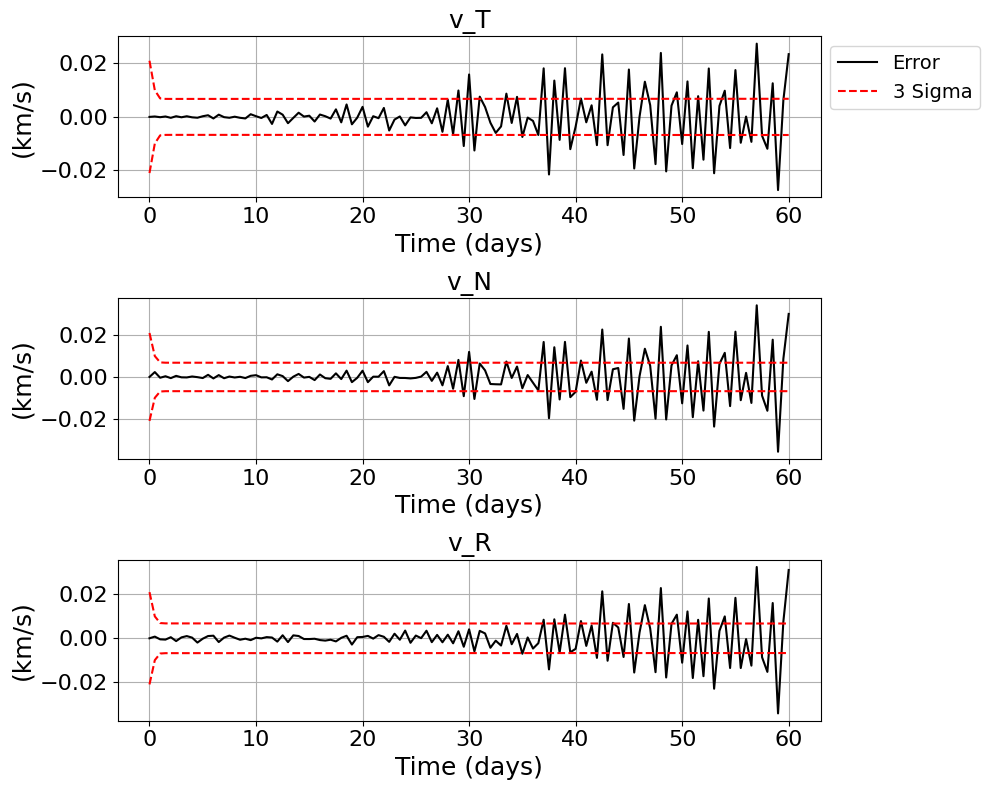}
    \caption{Velocity estimation error.}
\label{figsub:filter_pos_crab_J0030_J0437}  
  \end{subfigure}
  \caption{Navigation filter performance in the spacecraft RTN frame for a segment of the Earth-Jupiter transfer orbit when pulsar combination \#1 is used (Crab, J0030+0451 and J0437-4715). The filter starts to diverge after about 20 days, due to the large extrapolation error growth of the Crab's timing model.}
\label{fig:filter_crab_J0030_J0437}
\end{figure}

\begin{figure}[tbp!]
  \centering
  \begin{subfigure}[t]{0.49\textwidth}
    \centering
    \includegraphics[width=1\linewidth,trim={0 0 0 0.04in},clip]{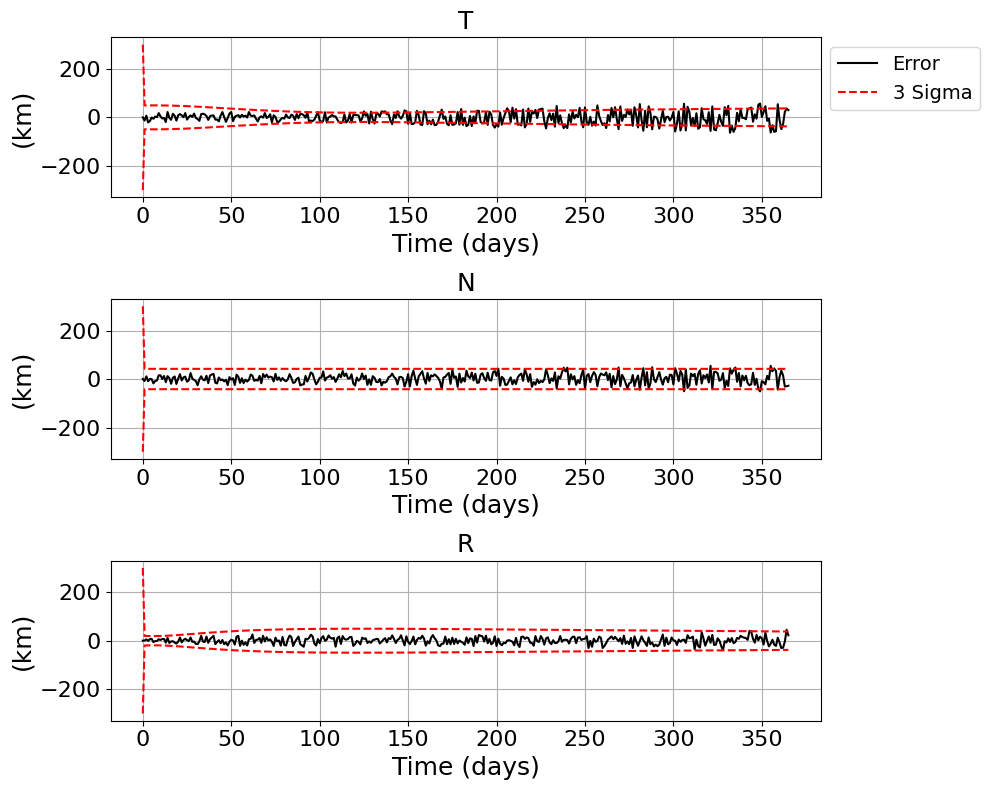}
    \caption{Position estimation error.}
\label{figsub:filter_pos_J2124_J0030_J0437}
  \end{subfigure}
  \begin{subfigure}[t]{0.49\textwidth}
    \centering
\includegraphics[width=1\linewidth]{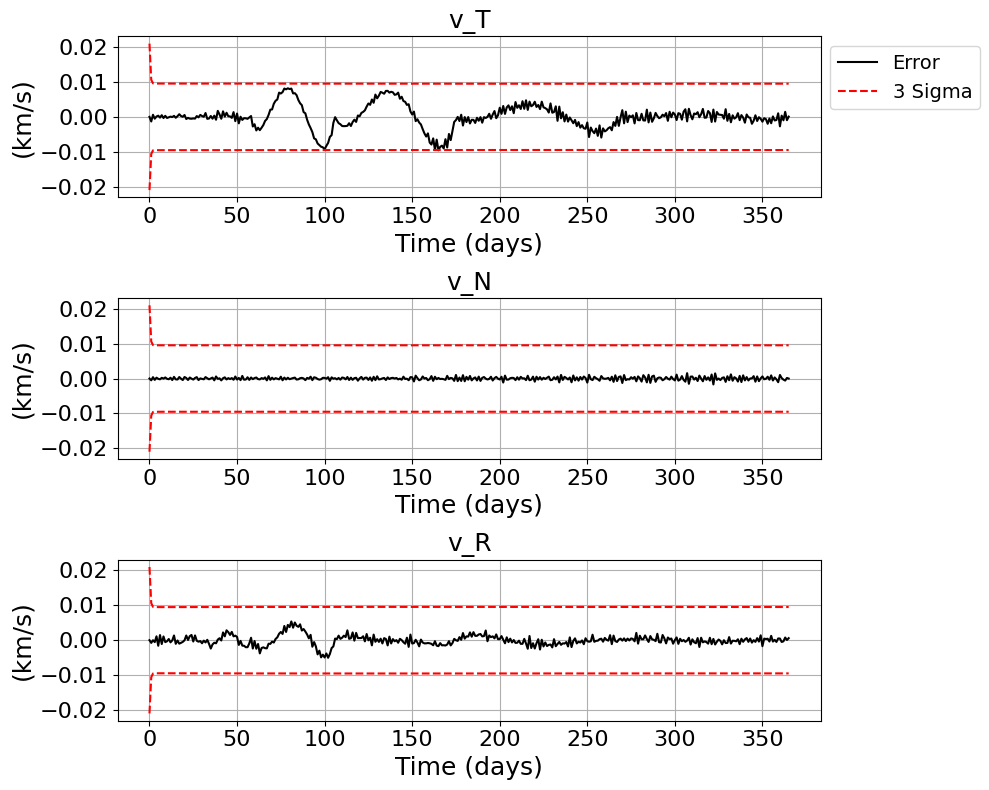}
    \caption{Velocity estimation error.}
\label{figsub:filter_vel_J2124_J0030_J0437}  
  \end{subfigure}
  \caption{Navigation filter performance in the spacecraft RTN frame for a segment of the Earth-Jupiter transfer orbit when pulsar combination \#2 is used (J2124-3358, J0030+0451, J0437-4715). The filter remains converged throughout the 365-day simulation.}
\label{fig:filter_J2124_J0030_J0437}
\end{figure}

\begin{figure}[tbp!]
  \centering
  \begin{subfigure}[t]{0.49\textwidth}
    \centering
    \includegraphics[width=1\linewidth,trim={0 0 0 0.04in},clip]{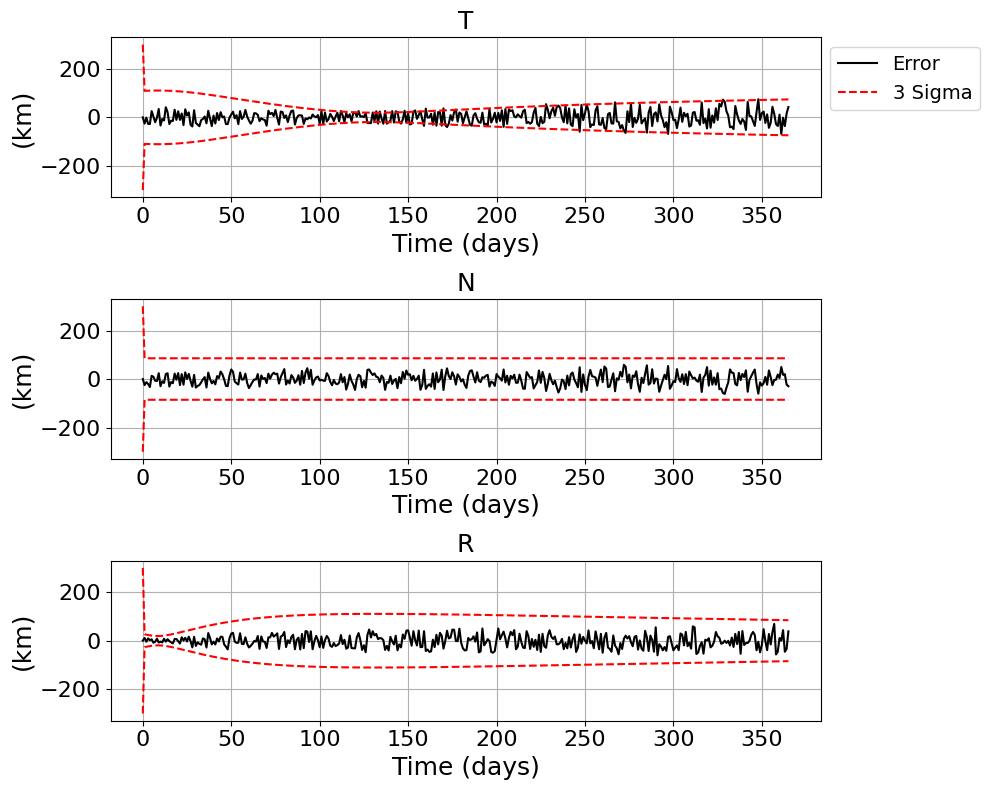}
    \caption{Position estimation error.}
\label{figsub:filter_pos_J1012_J0030_J0437}
  \end{subfigure}
  \begin{subfigure}[t]{0.49\textwidth}
    \centering
\includegraphics[width=1\linewidth]{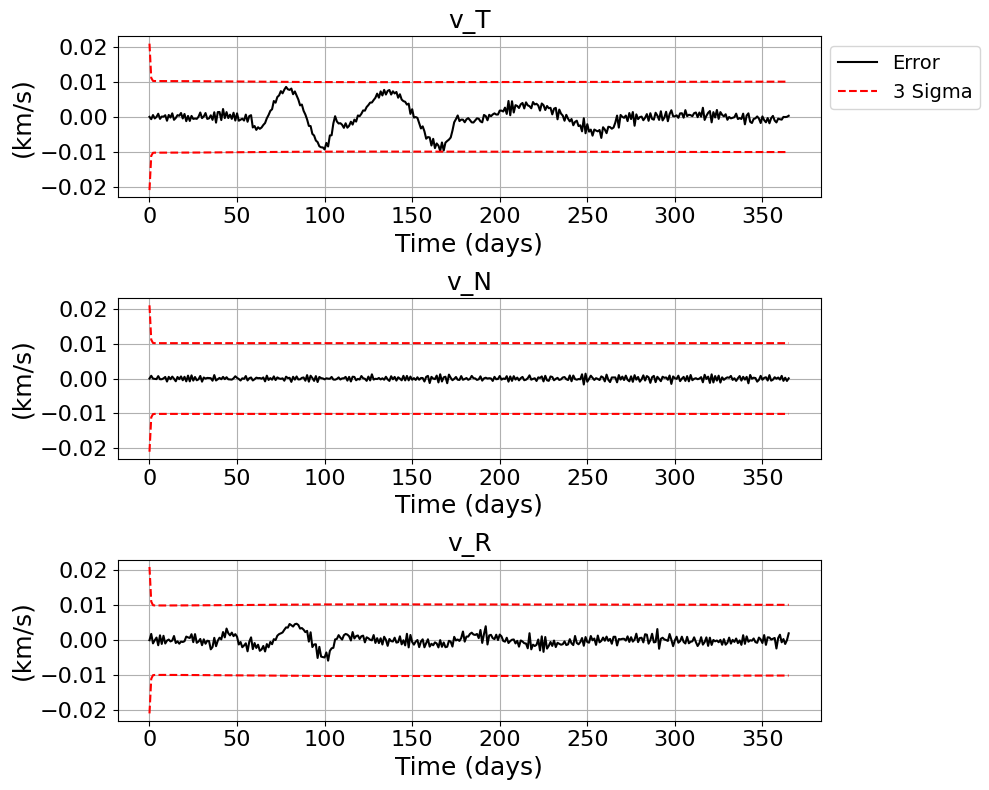}
    \caption{Velocity estimation error.}
\label{figsub:filter_vel_J1012_J0030_J0437}  
  \end{subfigure}
  \caption{Navigation filter performance in the spacecraft RTN frame for a segment of the Earth-Jupiter transfer orbit when pulsar combination \#3 is used (J1012+5307, J0030+0451, J0437-4715). The filter remains converged throughout the 365-day simulation.}
\label{fig:filter_J1012_J0030_J0437}
\end{figure}

Within the time frame considered, the navigation filter is run using each of these combinations. The filter performance is assessed by recording the position and velocity estimation errors in the spacecraft RTN frame. The corresponding time histories of the estimation errors are shown in Figures~\ref{fig:filter_crab_J0030_J0437}, \ref{fig:filter_J2124_J0030_J0437}, and \ref{fig:filter_J1012_J0030_J0437}, respectively. A quantitative summary of the steady-state filter 3$\sigma$ errors achieved with each combination is provided in Table~\ref{tab:filter_results}.

\begin{table*}[tb!]
\caption{Navigation filter performance (3$\sigma$ errors) for the Earth-Jupiter interplanetary spacecraft in the spacecraft RTN frame when different pulsar combinations are used.}
\centering
\resizebox{\textwidth}{!}{%
\begin{tabular}{cccc}
\toprule
 &
  \begin{tabular}[c]{@{}c@{}}Combination \#1\\ (Crab, J0030+0451, J0437-4715)\end{tabular} &
  \begin{tabular}[c]{@{}c@{}}Combination \#2\\ (J2124-3358, J0030+0451, J0437-4715)\end{tabular} &
  \begin{tabular}[c]{@{}c@{}}Combination \#3\\ (J1012+5307, J0030+0451, J0437-4715)\end{tabular} \\ 
  \midrule
$T$ (km)    & 15.96573    & 37.05783   & 74.00871   \\
$N$ (km)    & 18.41399    & 41.97826   & 85.55772   \\
$R$ (km)    & 8.39080    & 37.89489    & 84.47903   \\
$v_T$ (km/s) & 0.000299218 & 0.009517576 & 0.010097465 \\
$v_N$ (km/s) & 0.000311489 & 0.009529412 & 0.010146865 \\
$v_R$ (km/s) & 0.000200623 & 0.009519519 & 0.010141844 \\ \bottomrule
\end{tabular}
}
\label{tab:filter_results}
\end{table*}

From the results, Combination \#1 yields the highest navigation accuracy, with position errors below 20 km in all R-T-N directions and velocity errors under 0.5 m/s. This performance is primarily due to the inclusion of the Crab pulsar, the brightest source among all, which provides the strongest signal and leads to the highest accuracy for a given detector size. However, the filter performance degrades rapidly after approximately 20 days, with errors growing beyond the $3\sigma$ bounds.  This degradation is driven by the low timing stability of the Crab pulsar, which, being the youngest among those considered, suffers from rapidly growing phase prediction error when its timing model is extrapolated. As evidenced in Figure~\ref{figsub:extrapolate_model_error_1yr}, the Crab’s extrapolated phase error increases sharply, almost vertically, and quickly reaches a large value after 20 days, in contrast to the slower and more gradual error growth observed in the other pulsars. As a result, the measurement errors become too large for the filter to handle, causing the filter to diverge beyond 20 days. These results suggest that while the Crab pulsar can significantly improve navigation accuracy, maintaining filter convergence would require telemetry updates to its timing model approximately every 20 days.

In contrast, Combinations \#2 and \#3 do not include the Crab pulsar and result in larger estimation errors. However, the pulsars in these combinations exhibit more stable timing behaviour over the full simulation period, as shown in Figure~\ref{figsub:extrapolate_model_error_1yr}. Consequently, the filter remains stable, with estimation errors staying within the $3\sigma$ bounds for the entire 365-day simulation duration, without requiring any timing model updates. This illustrates a key trade-off: improved navigation accuracy with frequent timing model updates versus long-term autonomy with reduced accuracy.

\subsubsection{Case 2: Low Earth Orbit Satellite}

\begin{figure}[tbp!]
  \centering \includegraphics[width=0.6\linewidth]{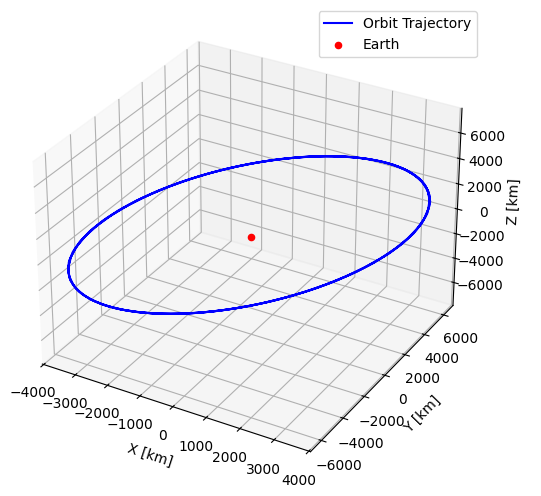}
  \caption{LEO circular orbit with an altitude of 600~km and an orbital period of about 1.6 hours.}
\label{fig:orbi_LEO}
\end{figure}

The navigation simulation is also run for the LEO scenario. The satellite is assumed to have the same properties as those listed in Table~\ref{tab:sc_properties}, while its initial orbital conditions and launch date are given in Table~\ref{tab:launch_conditions_LEO}. The orbital period is approximately 1.6 hours, and the resulting orbit is shown in Figure~\ref{fig:orbi_LEO}.

\begin{table}[!]
\caption{Launch date and the initial orbital conditions for a LEO satellite.}
\centering
\begin{tabular}{cccc}
\toprule
\multicolumn{2}{c}{Launch date}             & \multicolumn{2}{c}{01 Oct 2025} \\
\midrule
$x_0$ (km) & \multicolumn{1}{c}{3.520418e+03} & $vx_0$ (km/s)     & 3.513734e-01    \\
$y_0$ (km) & \multicolumn{1}{c}{5.938515e+03} & $vy_0$ (km/s)     & -1.466165e+00     \\
$z_0$ (km) & \multicolumn{1}{c}{1.007117e+03} & $vz_0$ (km/s)     & 7.406608e+00     \\ 
\bottomrule
\end{tabular}
\label{tab:launch_conditions_LEO}
\end{table}

To evaluate the filter performance, a 5-hour simulation is conducted. As in Case 1, pulsar combinations are selected by considering the criteria previously discussed, prioritising high photon flux and high timing stability. From Figure~\ref{figsub:extrapolate_model_error_1yr}, pulsars B1509-58, J0751+1807, and J1231-1411 are first excluded due to their large initial phase prediction errors. Additionally, pulsars are selected to provide balanced geometric coverage in all three directions, while avoiding configurations in which all selected pulsars become simultaneously unobservable during any portion of the orbit. Based on these considerations, three pulsar sets are selected for analysis. The resultant navigation performance with each set is then assessed and compared.

\begin{itemize}
    \item Combination \#1: Crab, B1937+21, J0030+0451
    \item Combination \#2: B1937+21, J0030+0451, J0437-4715
    \item Combination \#3: B1937+21, J0030+0451, J1012+5307
\end{itemize}

\begin{figure}[tbp!]
  \centering
  \begin{subfigure}[t]{0.49\textwidth}
    \centering
    \includegraphics[width=1\linewidth,trim={0 0 0 0.04in},clip]{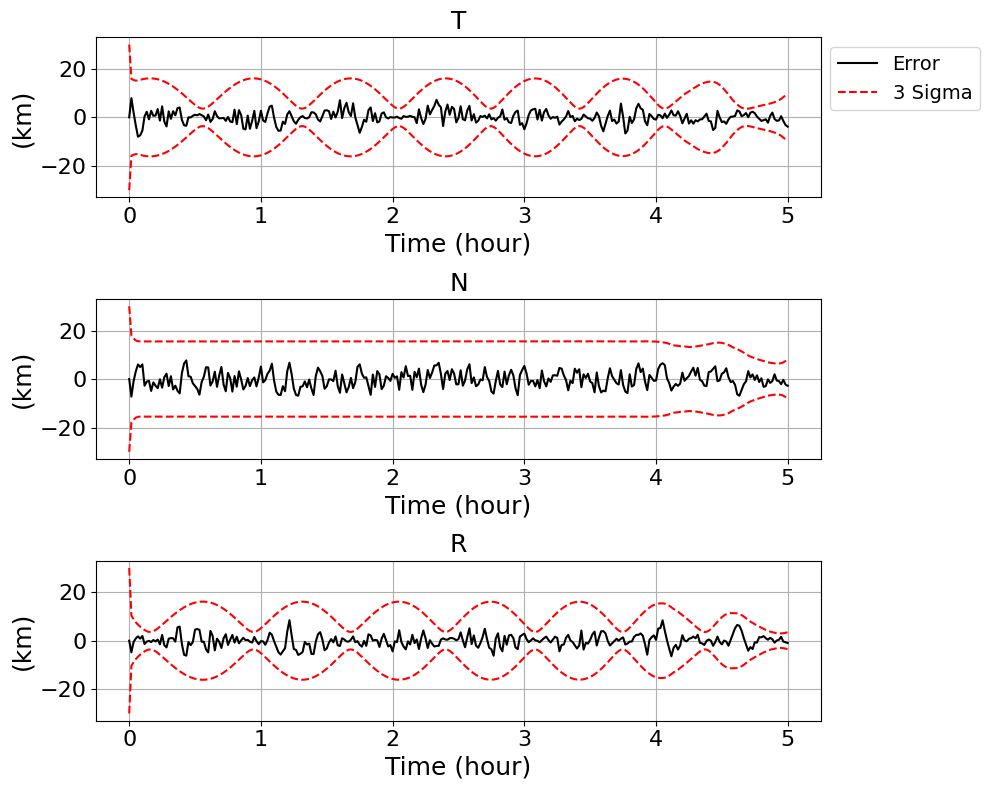}
    \caption{Position estimation error.}
\label{figsub:filter_LEO_pos_crab_B1937_J0030}
  \end{subfigure}
  \begin{subfigure}[t]{0.49\textwidth}
    \centering
\includegraphics[width=1\linewidth]{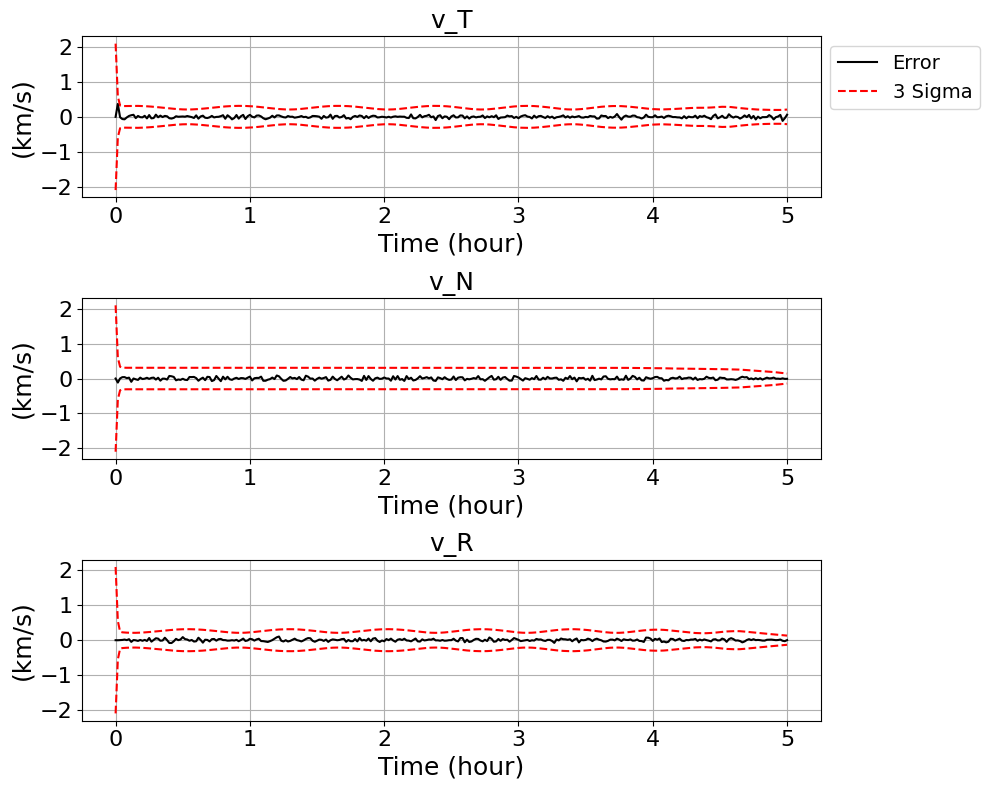}
    \caption{Velocity estimation error.}
\label{figsub:filter_LEO_vel_crab_B1937_J0030}  
  \end{subfigure}
  \caption{Navigation filter performance in the RTN frame for the LEO satellite when pulsar combination \#1 is used (Crab, B1937+21 and J0030+0451). The filter is run for 5 hours.}
\label{fig:filter_LEO_crab_B1937_J0030}
\end{figure}

\begin{figure}[tbp!]
  \centering
  \begin{subfigure}[t]{0.49\textwidth}
    \centering
    \includegraphics[width=1\linewidth,trim={0 0 0 0.04in},clip]{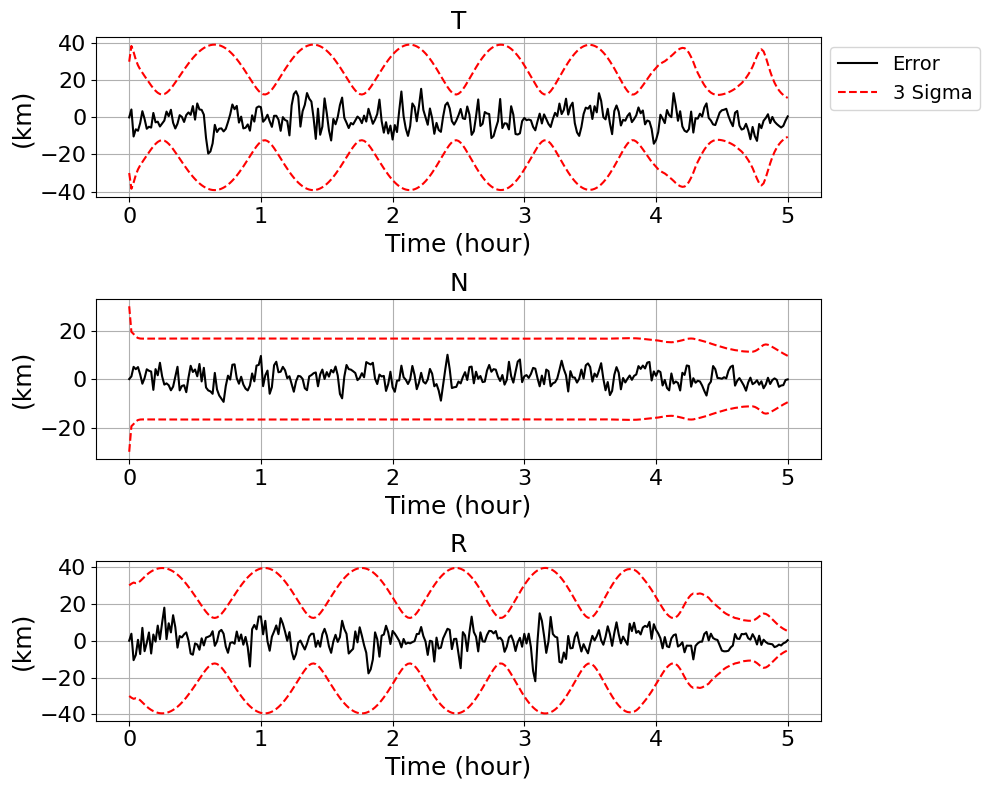}
    \caption{Position estimation error.}
\label{figsub:filter_LEO_pos_J0030_B1937_J0437}
  \end{subfigure}
  \begin{subfigure}[t]{0.49\textwidth}
    \centering
\includegraphics[width=1\linewidth]{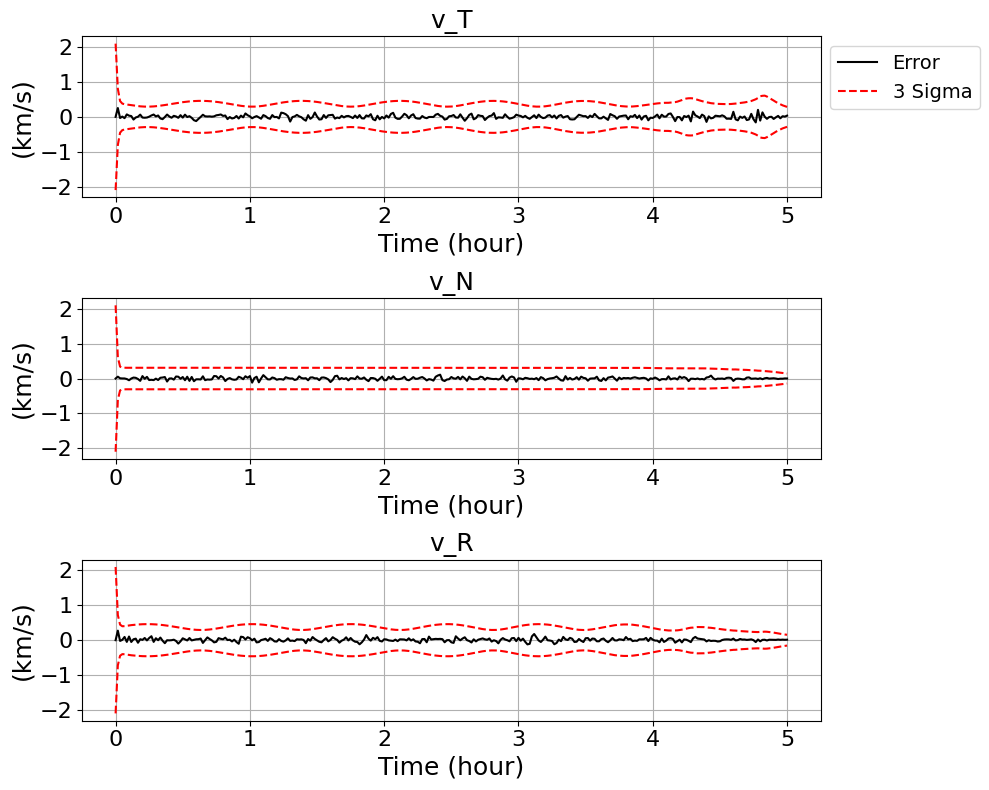}
    \caption{Velocity estimation error.}
\label{figsub:filter_LEO_vel_J0030_B1937_J0437}  
  \end{subfigure}
  \caption{Navigation filter performance in the RTN frame for the LEO satellite when pulsar combination \#2 is used (B1937+21, J0030+0451, J0437-4715). The filter is run for 5 hours.}
\label{fig:filter_LEO_J0030_B1937_J0437}
\end{figure}

\begin{figure}[tbp!]
  \centering
  \begin{subfigure}[t]{0.49\textwidth}
    \centering
    \includegraphics[width=1\linewidth,trim={0 0 0 0.04in},clip]{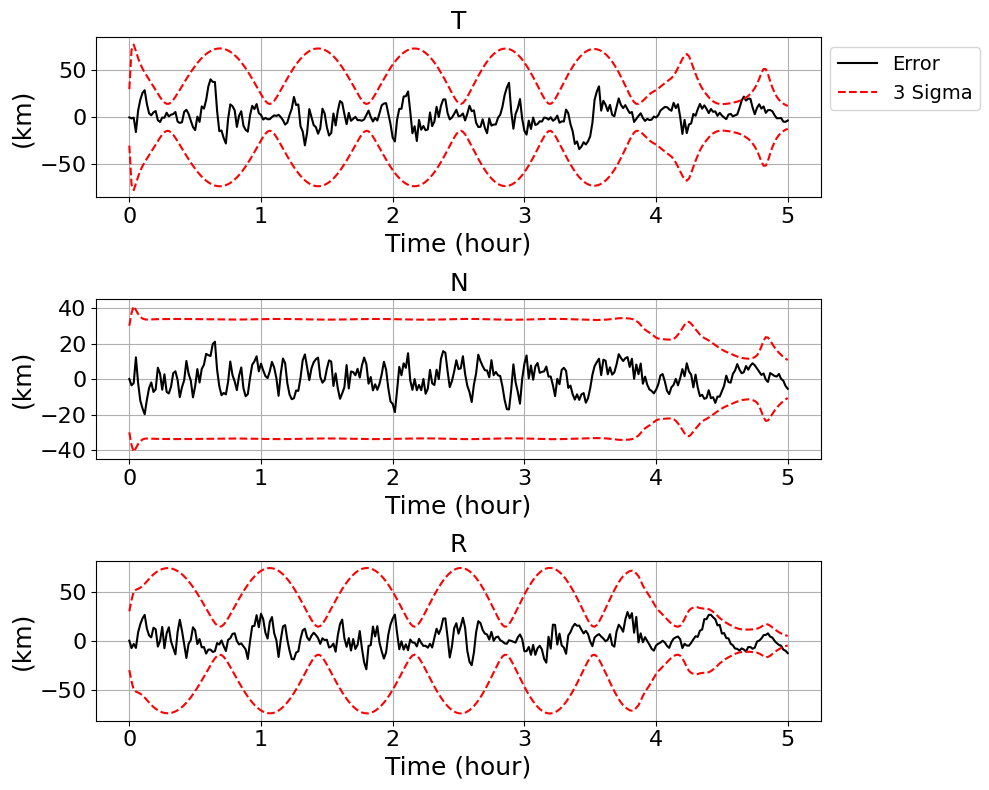}
    \caption{Position estimation error.}
\label{figsub:filter_LEO_pos_B1937_J0030_J1012}
  \end{subfigure}
  \begin{subfigure}[t]{0.49\textwidth}
    \centering
\includegraphics[width=1\linewidth]{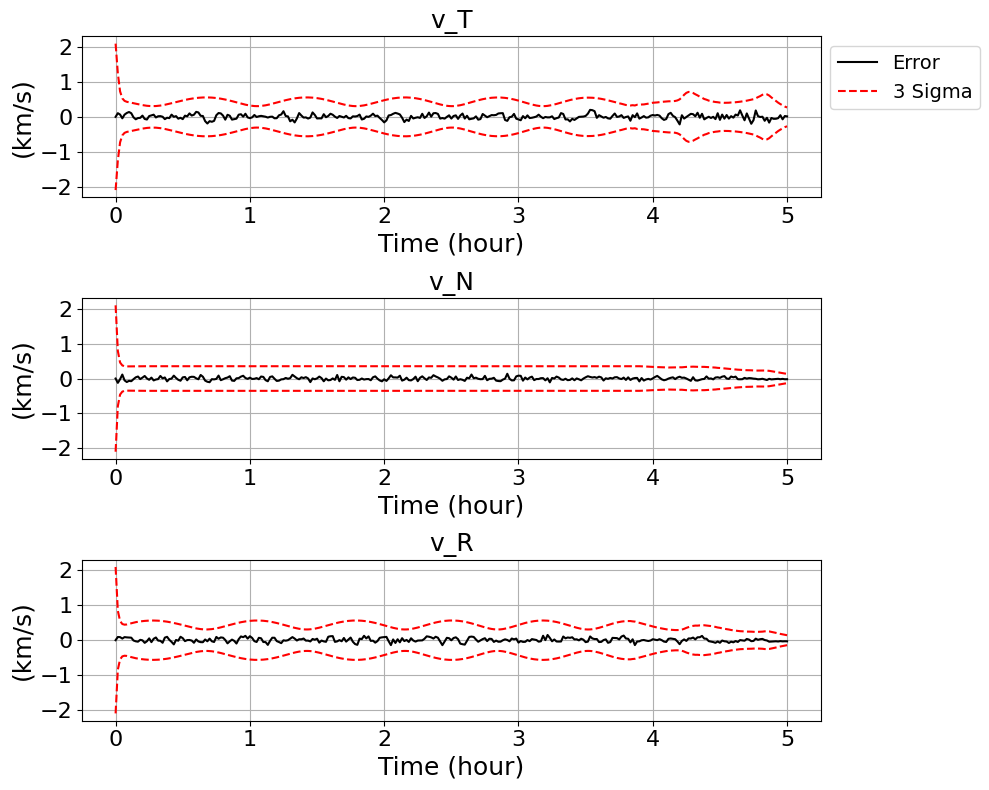}
    \caption{Velocity estimation error.}
\label{figsub:filter_LEO_vel_B1937_J0030_J1012}  
  \end{subfigure}
  \caption{Navigation filter performance in the RTN frame for the LEO satellite when pulsar combination \#3 is used (B1937+21, J0030+0451, J1012+5307). The filter is run for 5 hours.}
\label{fig:filter_LEO_B1937_J0030_J1012}
\end{figure}

The filter estimation errors in the RTN frame for each pulsar combination are shown in Figures~\ref{fig:filter_LEO_crab_B1937_J0030}, \ref{fig:filter_LEO_J0030_B1937_J0437}, and \ref{fig:filter_LEO_B1937_J0030_J1012}. The corresponding steady-state 3$\sigma$ accuracies are summarised in Table~\ref{tab:filter_results_LEO}

\begin{table*}[tb!]
\caption{Navigation filter performance (3$\sigma$ errors) for the LEO satellite in the satellite RTN frame when different pulsar combinations are used.}
\centering
\resizebox{\textwidth}{!}{%
\begin{tabular}{cccc}
\toprule
 &
  \begin{tabular}[c]{@{}c@{}}Combination \#1\\ (Crab, B1937+21, J0030+0451)\end{tabular} &
  \begin{tabular}[c]{@{}c@{}}Combination \#2\\ (B1937+21, J0030+0451, J0437-4715)\end{tabular} &
  \begin{tabular}[c]{@{}c@{}}Combination \#3\\ (B1937+21, J0030+0451, J1012+5307)\end{tabular} \\ 
  \midrule
$T$ (km)    & 3.53478   & 10.47792   & 12.05439   \\
$N$ (km)    & 6.39046   & 9.55427   & 10.65149   \\
$R$ (km)    & 3.00582    & 5.24670   & 4.84582  \\
$v_T$ (km/s) & 0.19693 & 0.28536 & 0.26923 \\
$v_N$ (km/s) & 0.14351 & 0.14385 & 0.13403 \\
$v_R$ (km/s) & 0.13176 & 0.15482 & 0.14262 \\ \bottomrule
\end{tabular}
}
\label{tab:filter_results_LEO}
\end{table*}

Combination \#1 includes pulsars with the highest photon fluxes. As a result, it achieves the best navigation accuracy, with position errors remaining below 7~km in all R-T-N directions. However, this set includes the Crab pulsar, whose poor long-term timing stability is known to cause filter divergence after approximately 20 days, as previously shown in Figure~\ref{fig:filter_crab_J0030_J0437}. Over the shorter simulation time span considered here (on the order of hours), this degradation does not arise and is therefore not directly visible in the plotted results.

To mitigate the need for frequent updates of the Crab’s timing model over longer timescales, Combinations \#2 and \#3 replace it with alternative pulsars, selected based on a trade-off between timing stability, visibility, and directional sensitivity. In Combination \#2, the Crab is replaced with J0437-4715, which has superior timing stability, as shown in Figure~\ref{fig:extrapolate_model_error}. In Combination \#3, the replacement is J1012+5307, chosen for its high visibility throughout the orbit, with only a short period of visibility loss (Figure~\ref{fig:visibility_EarthShadow_allpulsars_LEO600km}).

As summarised in Table~\ref{tab:filter_results_LEO}, both alternative pulsar combinations yield larger navigation errors across all three directions compared to Combination \#1. Nevertheless, they are expected to offer greater robustness for long-duration operations, either by reducing the frequency of timing model updates or by ensuring more consistent pulsar visibility along the orbit.

Overall, the results from this section highlight a key point in XNAV: the inherently multifaceted nature of pulsar selection for spacecraft navigation. Beyond photon flux, other factors, including timing stability, geometric configuration, and visibility, substantially affect navigation performance. While high flux may drive instantaneous accuracy, maintaining filter convergence over extended durations in space requires consideration of long-term timing predictability. Pulsar selection for spacecraft navigation must therefore balance competing objectives: minimising state estimation errors and minimising reliance on ground-based timing model updates.

It should be noted that the pulsar selection in this study was performed manually, based on a set of heuristic considerations of various criteria. A more generalised and mission-tailored pulsar selection strategy that optimises for visibility, geometric configuration, flux, and timing stability will be explored in future work. Developing such a generalised selection algorithm is a crucial next step towards enabling adaptable, scalable pulsar navigation systems across diverse mission scenarios.

It is also important to note that the navigation method presented in this paper is designed to refine an existing estimate of the spacecraft position, under the assumption that some prior knowledge of the position is available. This is different  from a lost-in-space scenario, where no initial position information is assumed. Nonetheless, by using the same pulsars and pulsar timing techniques, the approach can, in principle, be extended to address the lost-in-space case. This would, however, require different and more sophisticated pulsar signal processing algorithms.

\section{Conclusion}
\label{sec: conclusion}

This paper addressed key practical considerations for pulsar selection in the implementation of X-ray pulsar navigation for autonomous space missions. In contrast to previous studies that estimated measurement uncertainty using analytical models, this work directly uses re-scaled NICER observation data to evaluate achievable range accuracy and its dependence on pulsed flux, instrument effective area, and observation duration. The analysis incorporated a set of key mission-level constraints, including pulsed flux, pulsar visibility, geometric configuration, and long-term timing stability, that are critical for real-world pulsar selection but often overlooked in earlier studies.

An EKF is used as the navigation filter for spacecraft state estimation. Two scenarios are considered to test the performance of the XNAV system: an interplanetary transfer from Earth to Jupiter, and a LEO satellite in a circular orbit at 600 km altitude. Simulation results show that when the Crab pulsar is included, the position estimation error remains below 7 km in LEO and 20 km during interplanetary transfer with an instrument effective area of 200~cm$^2$. However, the Crab's poor timing stability leads to filter divergence after 20 days without timing model updates. In contrast, more stable pulsars enable long-term autonomy but with reduced accuracy. This highlights a key trade-off in pulsar selection. High-flux sources, such as the Crab pulsar, substantially improve short-term navigation accuracy, but may lead to filter divergence over time due to their poor long-term timing stability unless regular timing model updates are available. Conversely, more stable pulsars enable fully autonomous long-duration navigation, though with lower overall accuracy.

These findings highlight the multifaceted nature of pulsar selection for navigation and the need to balance competing objectives, in particular minimising estimation errors and maximising autonomy. To this end, future work will focus on developing a systematic, mission-tailored pulsar selection algorithm that jointly considers flux, visibility, geometry, and timing stability. This will be a key step towards enabling scalable and robust XNAV systems across diverse mission scenarios.

\section*{Declarations}

\subsection*{Availability of data and materials}
The datasets generated during the current study are available from the corresponding author on reasonable request.

\subsection*{Competing interests}
The authors declare that they have no competing interests.

\subsection*{Funding}

SC and FT would like to acknowledge the support received by the European Research Council (ERC) through the EXTREMA project under the European Union’s Horizon 2020 research and innovation programme (Grant agreement No. 864697; PI: Topputo). EP is supported by a Juan de la Cierva Fellowship (JDC2022-049957-I). NR and EP are supported by the ERC Consolidator Grant “MAGNESIA” (No. 817661; PI: Rea) and the Proof of Concept ``DeepSpacePulse" (No. 101189496; PI: Rea), by the Catalan grant SGR2021-01269 (PI: Graber/Rea), the Spanish grant ID2023-153099NA-I00 (PI: Coti Zelati), and by the program Unidad de Excelencia Maria de Maeztu CEX2020-001058-M. SC has received additional support by the ERC Proof of Concept grant ``DeepSpacePulse" (No. 101189496; PI: Rea). 

\subsection*{Authors' contributions}
SC: Methodology, Software, Data Collection and Analysis, Validation, Writing. EP: Software, Data Collection, Writing. NR: Conceptualisation, Supervision,  Funding acquisition. FT: Supervision, Project Administration, Funding Acquisition. All authors have reviewed and approved the final manuscript.

\subsection*{Acknowledgements}
We thank Serni Rib\'o for the fruitful interactions that contributed to this work. NR thanks Paul Ray for enlightening discussions on pulsar-based navigation, and all contributors to the PODIUM project (SENER, Deimos, and IEEC; Cacciatore et al. 2023).












\end{document}